\documentclass[12pt]{article}

\usepackage{amsmath}
\usepackage{amssymb}
\usepackage{cite}
\usepackage{ulem}

\usepackage{lineno}

\usepackage{microtype}
\DisableLigatures[f]{encoding = *, family = * }

\usepackage{graphics}
\usepackage{graphicx}
\usepackage{color}
\usepackage{here}
\usepackage[hyphens]{url}
\usepackage{hyperref}
\usepackage[hyphenbreaks]{breakurl}

\usepackage[labelfont=bf,labelsep=period,justification=raggedright]{caption}

\makeatletter
\renewcommand{\@biblabel}[1]{\quad#1.}
\makeatother

\date{}

\begin{document}

\begin{flushleft}
{\Large
\textbf{Effects of the distant population density on spatial patterns of demographic dynamics}
}
\\
Kohei Tamura$^{1, 2}$, 
Naoki Masuda$^{1}$
\\
\bf{1} Department of Engineering Mathematics, University of Bristol, Merchant Venturers Building, Woodland Road, Clifton, Bristol BS8 1UB, UK\\
\bf{2} Frontier Research Institute for Interdisciplinary Sciences, Tohoku University, Aramaki aza Aoba 6-3, Aoba-ku, Sendai 980-8578, Japan
\end{flushleft}

\subsection*{Author for correspondence:}
Naoki Masuda\\
e-mail: naoki.masuda@bristol.ac.uk

\section*{Abstract}
Spatiotemporal patterns of population changes within and across countries have various implications. Different geographical, demographic and econo-societal factors seem to contribute to migratory decisions made by individual inhabitants. Focussing on internal (i.e., domestic) migration, we ask whether individuals may take into account the information on the population density in distant locations to make migratory decisions. We analyse population census data in Japan recorded with a high spatial resolution (i.e., cells of size 500 m $\times$ 500 m) for the entirety of the country and simulate demographic dynamics induced by the gravity model and its variants. We show that, in the census data, the population growth rate in a cell is positively correlated with the population density in nearby cells up to a \textcolor{black}{distance} of 20 km as well as that of the focal cell. The ordinary gravity model does not capture this empirical observation. We then show that the empirical observation is \textcolor{black}{better} accounted for by extensions of the gravity model such that individuals are assumed to perceive the attractiveness, approximated by the population density, of the source or destination cell of migration as the spatial average over a \textcolor{black}{circle of} radius $\approx 1$ km. 

\section*{Keywords}
demography, dynamics, gravity model, migration, population census

\newpage
\section*{Introduction}
Demography, particularly spatial patterns of population changes, has been a target of intensive research because of its economical and societal implications, such as difficulties in upkeep of infrastructure \cite{Pallagst2007, Pallagst2013, Hara2014}, policy making related to city planning \cite{Pallagst2007, Pallagst2013} and integration of municipalities \cite{Hara2014}.
A key factor shaping spatial patterns of demographic dynamics is migration.
Migration decisions by inhabitants are affected by various factors including job opportunities, cost of living and climatic conditions \cite{Lewis1982, Kahley1991, Beine2014}. 
These and other factors are often non-randomly distributed in space, creating spatial patterns of migration and population changes over time.
A number of models have been proposed to describe and predict spatiotemporal patterns of human migration \cite{Stouffer1940, Zipf1946, Cohen2008, Batty2013, Simini2012, Simini2013, Barthelemy2016}.

Among these models, a widely used model is the gravity model (GM) and its variants \cite{Zipf1946, Anderson2011, Batty2013, Rodrigue2013}.
The GM assumes that the migration flow from one location to another is proportional to a power (or a different monotonic function) of the population at the source and destination locations and the distance between them.
The model has attained reasonably accurate description of human migration in some cases \cite{Zipf1946, Karemera2000, Fagiolo2013}, as well as other phenomena such as international trades \cite{Bhattacharya2008, Kepaptsoglou2010} and the volume of phone calls between cities \cite{Lambiotte2008, Krings2009}.

Studies of migration, such as those using the GM\cite{Zipf1946, Fagiolo2013} and other migration models \cite{Simini2012, Akwawua2016}, are often based on subdivisions of the space that define the unit of analysis such as administrative units (e.g., country and city). 
However, the choice of the unit of analysis is often arbitrary. 
Humans whose migratory behaviour is to be modelled microscopically, statistically or otherwise, may pay less attention to such a unit than a model assumes when they make a decision to move home. This may be particularly so for internal (i.e., domestic) migrations rather than for international migrations because boundaries of administrative units may less impact inhabitants in the case of internal migrations than international migrations. 
This issue is related to the modifiable areal unit problem in geography, which stipulates that different units of analysis may provide different results \cite{Openshaw1984}.
For example, particular partitions of geographical areas can affect parameter estimates of gravity models \cite{Openshaw1977}.
To overcome such a problem, criteria for selecting appropriate units of analysis have been sought \cite{Broadbent1970, Batty1974, Batty1976, Masser1977, Openshaw1977}.
Another strategy to address the issue of the unit of analysis is to employ models with a maximally high spatial resolution.
For example, a recently proposed continuous-space GM assumes that the unit of analysis is an infinitesimally small spatial segment \cite{Simini2013}. 
This approach implicitly assumes that the unit of analysis, which a modelled individual perceives, is an infinitesimally small spatial segment. 
In fact, humans may regard a certain spatial region, which may be different from an administrative unit and have a certain finite but unknown size, as a spatial unit based on which they make a migration decision.
If this is the case, individuals may make decisions by taking into account the environment in a neighbourhood of the current residence and/or the destination of the migration up to a certain \textcolor{black}{distance}. 
Here we examine this possibility by combining data analysis and modelling, complementing past research on the choice of geographical units for understanding human migration \cite{Broadbent1970, Batty1974, Batty1976, Masser1977, Openshaw1977}.

In this article, we analyse demographic data obtained from the population census of Japan carried out in 2005 and 2010, which are provided with a high spatial resolution \cite{estatjp}. 
We hypothesise that the growth rate of the population is influenced by the population density near the current location as well as that at the focal location, where each location is defined by a 500 m $\times$ 500 m cell in the grid according to which the data are organised. We provide evidence in favour of this hypothesis through correlation-based data analysis. Then, we argue that the GM is insufficient to produce the empirically observed spatial patterns of the population growth. We provide extensions of the GM that better fit the empirical data, in which individuals are assumed to aggregate the population of nearby cells to calculate the attractiveness of the source or destination cell of migration.

\section*{Methods}
\subsection*{Data set}
We analysed demographic dynamics using data from the population census in Japan \cite{estatjp}, which consisted of measurements from $K = 1,944,711$ cells of size 500 m $\times$ 500 m. 
The census is conducted every five years.
We used data from the censuses conducted in 2005 and 2010 because data with such a high spatial resolution over the entirety of Japan were only available for these years.
We also ran the following analysis using the data from the census conducted in 2000 (Appendix A), which were somewhat less accurate in counting the number of inhabitants in each cell than the data in 2005 and 2010 \cite{gaiyo}.
In the main text, we refer to the two time points 2005 and 2010 as $t_{1}$ and $t_{2}$, respectively.
The number of inhabitants in cell $i$ ($1\le i\le K$) at time $t$ is denoted by $n_{i}(t)$.
We used the latitude and longitude of the centroid of each cell to define its position.
Basic statistics of the data at the three time points are shown in Table \ref{tab:summary}.

\subsection*{Spatial correlation}
\textcolor{black}{We defined the distance between cells $i$ and $j$, denoted by $d_{ij}$, as that between the centroids of the two cells in kilometers.}
We measured the spatial correlation in the number of inhabitants between a pair of cells at distance $d$ by \cite{Gallos2012}
\begin{equation}
C(d) = \frac{1}{\sigma^2} \frac{\sum_{i^{\prime}=1}^{K^{\prime}} \sum_{j^{\prime}=1}^{K^{\prime}} (n_{i^{\prime}}-\overline{n})(n_{j^{\prime}}-\overline{n})I(d<d_{i^{\prime} j^{\prime}} \leq d+1)}{\sum_{i^{\prime}=1}^{K^{\prime}} \sum_{j^{\prime}=1}^{K^{\prime}} I(d<d_{i^{\prime}j^{\prime}} \leq d+1)},
\label{eq:spatial_cor}
\end{equation}
\textcolor{black}{which is essentially the Pearson correlation coefficient calculated from all pairs of cells  distance $\approx d$ apart. In Eq.~(1),} $\overline{n}=\sum_{i^{\prime}}^{K^{\prime}} n_{i^{\prime}}/K^{\prime}$ is the average number of inhabitants in an inhabited cell; $\sigma^2 = \sum_{i}^{K^{\prime}} (n_{i}-\overline{n})^2/K^{\prime}$ is the variance of the number of inhabitants in an inhabited cell; $I(d<d_{i^{\prime} j^{\prime}} \leq d+1)=1$ if $d<d_{i^{\prime} j^{\prime}} \leq d+1$ ($d=0,1, 2, \ldots$) and $I(d<d_{i^{\prime} j^{\prime}} \leq d+1)=0$ otherwise; $K^{\prime}$ (= 482,181 \textcolor{black}{at time $t_{1}$ and 477,172 at time $t_{2}$}) is the number of inhabited cells. In Eq.~(1), the summations on the right-hand side are restricted to the inhabited cells $i^{\prime}$ and $j^{\prime}$.
We suppressed the time in Eq.~(1).
It should be noted that $C(d)$ can be larger than 1.

\subsection*{Correlation between the growth rate and the population density in nearby cells}
\textcolor{black}{In the analysis of the growth rate of cells \textcolor{black}{described in this section}, we only used \textcolor{black}{focal} cells \textcolor{black}{$i$} whose population size was between 10 and 100 at $t_1$. 
We did so because the growth rate of less populated cells tended to fluctuate considerably and the growth rate of a more populated cell tended to be $\approx 0$. 
We carried out the same set of analysis for cells whose population size was greater than 100 to confirm that the main results shown in the following sections remain qualitatively the same (Appendix C).
It should be noted that cell $i$ may be partially water-surfaced.}

To calculate the correlation between the rate of population growth in a cell and the population density in cells nearby, we first divided the entire map of Japan into square regions of approximately 50 km $\times$ 50 km. 
The regions were tiled in a $64 \times 45$ grid to cover the entire Japan.
The minimum and maximum longitudes in the data set were 122.94 and 153.98, respectively. Therefore, we divided the range of the longitude into 64 windows, i.e., [122.4, 123), [123, 123.5), ..., [153.5, 154].
Similarly, the minimum and maximum latitudes were 45.5229 and 24.0604, respectively.
We thus divided the range of the latitude into 45 windows, i.e., [24, 24,5), [24.5, 25), ..., [45.5,46].
We classified each cell into one of the 64 $\times$ 45 regions on the basis of the coordinate of the centroid of the cell.
Note that there were sea regions without any inhabitant.
A region included \textcolor{black}{9,600} cells at most.

The growth rate of cell $i$ in the five years is given by
\begin{equation}
R_{i}=\frac{n_{i}(t_2)-n_{i}(t_1)}{n_{i}(t_1)}.
\end{equation}
We denoted by $D_{i}(\textcolor{black}{d})$ the population density at time $t_1$ averaged over the cells \textcolor{black}{$j$} whose distance from cell $i$, $\textcolor{black}{d}_{ij}$, is approximately equal to $\textcolor{black}{d}$, i.e., $\textcolor{black}{d}<\textcolor{black}{d}_{ij} \leq \textcolor{black}{d}+1$. 
We calculated the Pearson correlation coefficient between the population growth rate \textcolor{black}{(i.e., $R_i$)} and $D_i(\textcolor{black}{d})$, restricted to \textcolor{black}{the cells in} region $k$, i.e.,

\begin{equation}
\rho_{k}(\textcolor{black}{d}) = \frac{\sum_{i=1\text{; cell } i \in \text{region } k}^K (R_{i}-\overline{R}_{k})(D_{i}(\textcolor{black}{d})-\overline{D}_{k}(\textcolor{black}{d}))}{\sqrt{ \sum_{i=1\text{; cell } i \in \text{region } k}^K (R_{i} - \overline{R}_{k})^2} \sqrt{ \sum_{i=1\text{; cell } i \in \text{region } k}^K (D_{i}(\textcolor{black}{d}) - \overline{D}_{k}(\textcolor{black}{d}))^2 }},
\label{eq:rho}
\end{equation}
where $\overline{R}_{k}$ and $\overline{D}_{k}(\textcolor{black}{d})$ are the average of $R_{i}$ and $D_{i}(\textcolor{black}{d})$ over the cells in region $k$, respectively. 
\textcolor{black}{A positive value of $\rho_{k}(\textcolor{black}{d})$ is consistent with our hypothesis that the population growth rate is influenced by the population density in different cells.}
\textcolor{black}{We remind that the summation in Eq.~(3) is taken over the cells whose population is between 10 and 100.}
The correlation coefficient $\rho_{k}(\textcolor{black}{d})$ ranges between $-1$ and $1$.
\textcolor{black}{We did not exclude water-surface cells or partially water-surface cells $j$ from the calculation of $D_i(d)$.}
Finally, we defined $\overline{\rho}(\textcolor{black}{d})$ as the average of $\rho_{k}(\textcolor{black}{d})$ over all regions excluding those with less than 20 populated cells.
We decided to calculate $\rho_{k}(\textcolor{black}{d})$ for individual regions, $k$, and averaged it over the regions rather than to calculate the single correlation coefficient between $R_i$ and $D_i(\textcolor{black}{d})$ for the entirety of Japan. In this way, we aimed to suppress fluctuations in individual $\rho_{k}(\textcolor{black}{d})$.
We show $\rho_{k}(\textcolor{black}{d})$ for each region in Appendix B.
\textcolor{black}{We also show $\rho_{k}(d)$ for region $k$ such that all cells within region $k$ and those within 30 km from any cell in region $k$ are not in the sea in Appendix B.}

To examine the statistical significance of $\overline{\rho}(\textcolor{black}{d})$, we carried out bootstrap tests by shuffling the number of inhabitants in the populated cells at $t_{2}$ without shuffling that at $t_{1}$ and calculating $\overline{\rho}(\textcolor{black}{d})$.
We generated 100 randomized samples and calculated the distribution of $\overline{\rho}(\textcolor{black}{d})$ for each sample.
We deemed the value of $\overline{\rho}(\textcolor{black}{d})$ for the original data to be significant if it was not included in the $95 \%$ confidential interval (CI) calculated on the basis of the 100 randomized samples.

\subsection*{Gravity model}
In the standard gravity model (GM), the migration flow from source cell $i$ to destination cell $j$ ($\neq i$), $T_{ij}$, is given by
\begin{equation}\label{eq_gm}
T_{ij} = G\frac{n_{i}^{\alpha} n_{j}^{\beta}}{d_{ij}^{\gamma}},
\end{equation}
where $G$, $\alpha$, $\beta$ and $\gamma$ are parameters.
\textcolor{black}{Because $\alpha$, $\beta$, and $\gamma$ are usually assumed to be positive, Eq.~(4) implies that the migration flow is large when the source or the destination cell has many inhabitants or when the two cells are close to each other.}

In addition to the GM, we investigated two extensions of the GM in which the migration flow depends on the numbers of inhabitants in a neighborhood of cell $i$ or $j$.
The first extension, which we refer to as the GM with the spatially aggregated population density at the destination (d-aggregate GM), is given by
\begin{equation}\label{eq:d-aggregate}
T_{ij} = G\frac{n_{i}^{\alpha} N_{j}(\textcolor{black}{d}_{\rm ag})^{\beta}}{d_{ij}^{\gamma}},
\end{equation}
where $N_{j}(\textcolor{black}{d}_{\rm ag})$ is the number of inhabitants contained \textcolor{black}{in the cells within distance} $\textcolor{black}{d}_{\rm ag}$ km \textcolor{black}{from} cell $j$. 
\textcolor{black}{We remind that the distance between two cells is defined as that between the centroids of the two cells.}
The rationale behind this extension and the next one is that humans may perceive the population density at the source or destination as a spatial average.
\textcolor{black}{A similar assumption was used in a model of city growth, where cells close to inhabitant cells were more likely to be inhabited \cite{Rybski2013}.}

The second extension of the GM aggregates the population density around the source \textcolor{black}{cell}.
\textcolor{black}{To derive this variant of the GM, we rewrite Eq.~(4)} as $T_{ij} = n_{i}\times n_{i}^{\alpha-1} n_{j}^{\beta}/d_{ij}^{\gamma}$ \textcolor{black}{and interpret that each individual in cell $i$ is subject to the rate of moving to cell $j$, i.e., $n_i^{\alpha-1} n_j^{\beta} / d_{ij}^{\gamma}$}.
\textcolor{black}{The second extension, which we refer to as the GM with the aggregated population density at the source (s-aggregate GM), is defined by}
\begin{equation}\label{eq:s-aggregate}
T_{ij} = G n_{i} \frac{N_{i}(\textcolor{black}{d}_{\rm ag})^{\alpha-1} n_{j}^{\beta}}{d_{ij}^{\gamma}}.
\end{equation}
Unless we state otherwise, we set $\textcolor{black}{d}_{\rm ag} = 0.65$ in the d-aggregate and s-aggregate GMs, which is equivalent to the aggregation of a cell with the neighbouring four cells in the north, south, east, and west.
\textcolor{black}{We will also examine larger $d_{\rm ag}$ values.}

Using one of the three GMs, we projected the number of inhabitants in each cell at time $t_2$ given the empirical data at time \textcolor{black}{$t_1$}.
The predicted number of inhabitants in cell $i$ at time $t_2$, denoted by $\hat{n}_{i}(t_2)$, is given by
\begin{equation}
\hat{n}_{i}(t_2) = n_{i}(t_1) + \sum_{j=1}^{K} T_{ji} - \sum_{j=1}^{K} T_{ij}.
\end{equation}
We refer to $\sum_{j=1}^{K} T_{ji}$, $\sum_{i=1}^{K} T_{ij}$ and $\sum_{j=1}^{K} T_{ji} - \sum_{j=1}^{N} T_{ij}$ as the in-flow, out-flow and net flow of the population at cell $i$, respectively.

The projection of the growth rate, denoted by $\hat{R}_{i}$, is defined by $\hat{R}_{i} = \left[\hat{n}_{i}(t_2)-n_{i}(t_1)\right]/n_{i}(t_1) = \left(\sum_{j=1}^{K} T_{ji} - \sum_{j=1}^{K} T_{ij}\right)/n_{i}(t_1) $, based on which we calculated $\overline{\rho}(\textcolor{black}{d})$ for the model. We set $G=1$ because the value of $\overline{\rho}(\textcolor{black}{d})$ does not depend on $G$.

We measured the discrepancy between the empirical and projected data in terms of $\overline{\rho}(\textcolor{black}{d})$ by
\begin{equation}
\frac{\sum^{99}_{\textcolor{black}{d}=0} |\overline{\rho}_{\rm empirical}(\textcolor{black}{d})-\overline{\rho}_{\rm model}(\textcolor{black}{d})|}{\sum^{99}_{\textcolor{black}{d}=0}|\overline{\rho}_{\rm empirical}(\textcolor{black}{d})|},
\label{accuracy}
\end{equation}
where $\overline{\rho}_{\rm empirical}(\textcolor{black}{d})$ and $\overline{\rho}_{\rm model}(\textcolor{black}{d})$ are the values of $\overline{\rho}(\textcolor{black}{d})$ obtained \textcolor{black}{for} the empirical data and a model, respectively. 
\textcolor{black}{If the relationship between $\rho(\textcolor{black}{d})$ and $\textcolor{black}{d}$ is similar between the empirical data and the model, the discrepancy given by Eq.~(8) takes a small value.}

\section*{Results}
\subsection*{Spatial distribution of \textcolor{black}{inhabitants}}
\textcolor{black}{The} spatial distribution of \textcolor{black}{the number of inhabitants} at time $t_2$ is shown in Fig.~\ref{map}. 
The figure suggests centralization of the number of inhabitants in urban areas. 
We calculated the Gini index\textcolor{black}{, defined by $1/2K^{'2} \times \sum_{i^{'}=1}^{K^{'}}\sum_{j^{'}=1}^{K^{'}}|n_{i^{'}}-n_{j^{'}}|/\overline{n}$,} to quantify heterogeneity in the population density across cells\textcolor{black}{; it is often used for measuring wealth inequality}.
The Gini \textcolor{black}{index} at $t_{1}$ and $t_{2}$ \textcolor{black}{was} equal to 0.797 and 0.804, respectively, suggesting a high degree of \textcolor{black}{heterogeneity}.
\textcolor{black}{The} survival function of the number of inhabitants in a cell at $t_{1}$ and $t_{2}$ \textcolor{black}{is shown in Fig.~\ref{pop_dist}}.
\textcolor{black}{The figure} suggests that a majority of cells contains a relatively small number of inhabitants, whereas a small fraction of cells has many inhabitants.

Figure~\ref{map} \textcolor{black}{suggests the presence of} spatial correlation in the population \textcolor{black}{density}, as \textcolor{black}{observed} in other countries \cite{Gallos2012}.
\textcolor{black}{Therefore, we measured the} spatial correlation \textcolor{black}{coefficient} in the population size between a pair of cells, \textcolor{black}{$C(d)$, where $d$ was the distance between a pair of cells. Figure~\ref{spatial_cor}} indicat\textcolor{black}{es that $C(d)$ is substantially positive up to $d \approx 70$ km, confirming the presence of }spatial correlation.
This \textcolor{black}{correlation} length was shorter than \textcolor{black}{that observed} in previous studies \textcolor{black}{of data recorded} in the United States \cite{Gallos2012} ($\approx$ 1,000 km) and \textcolor{black}{spatial correlation in} the \textcolor{black}{population} growth rate in Spain \cite{Hernando2014} ($\approx$ 500 km) and the United States \cite{Hernando2015} (over 5,000 km). 

\textcolor{black}{\subsection*{Effects of the population density in nearby cells on migration}}
\textcolor{black}{We measured $\overline{\rho}(\textcolor{black}{d})$, which quantifies the effect of the population in cells at distance $\textcolor{black}{d}$ on the population growth in a focal cell.}
Figure~\ref{rho} shows $\overline{\rho}(\textcolor{black}{d})$ \textcolor{black}{as a function of $\textcolor{black}{d}$}.
The values of $\overline{\rho}(\textcolor{black}{d})$ were \textcolor{black}{the} largest at $\textcolor{black}{d}=0$.
\textcolor{black}{In other words, the effects of} the population density within 1 km \textcolor{black}{is the most positively correlated with the growth rate of a cell. This result reflect\textcolor{black}{s} the observation that highly populated cells tend to grow and vice versa \cite{Michaels2012, Birchenall2016, Desmet2015} (but see \cite{Rozenfeld2008})}.
As $\textcolor{black}{d}$ increased, $\overline{\rho}(\textcolor{black}{d})$ decreased and \textcolor{black}{reached $\approx 0$ for} $\textcolor{black}{d} \ge 20$ km.
This result suggests that cells surrounded by cells with a large (small) population density \textcolor{black}{within $\approx 20$ km} are more likely to \textcolor{black}{gain (lose) inhabitants}.

\textcolor{black}{The observed correlation between the population growth rate of a cell and the population of nearby cells may be explained by the combination of spatial correlation in the population density (Fig. 3) and positive correlation between the population growth rate and the population density in the same cell. To exclude this possibility, we measured $\overline{\rho}(\textcolor{black}{d})$ as the partial correlation coefficient, modifying Eq.~(\ref{eq:rho}), controlling for the population size of a focal cell. The results were qualitative the same as those based on the Pearson correlation coefficient (Appendix D).}

\subsection*{Gravity models}
Various mechanisms \textcolor{black}{may generate the dependence of the population growth rate in a cell on different cells (up to $\approx 20$ km apart), including heterogeneous birth and death rates that are spatially correlated. Here we} focussed on the effects of migration \textcolor{black}{as a possible mechanism to generate such a dependency}.
\textcolor{black}{We simulated migration dynamics using the gravity model \cite{Zipf1946, Batty2013, Rodrigue2013} and its variants and compared the projection obtained from the models with the empirical data}.
\textcolor{black}{We did not consider the radiation models \cite{Simini2012, Simini2013} including intervening opportunity models \cite{Stouffer1940} because our aim here was to qualitatively understand some key factors that may explain the effects of distant cells observed in Fig.~\ref{rho} rather than to reveal physical laws governing migration.}

In Fig.~\ref{rho}, we \textcolor{black}{compare} $\overline{\rho}(\textcolor{black}{d})$ \textcolor{black}{between} the empirical \textcolor{black}{data and those generated by} the GM, d-aggregate GM and s-aggregate GM.
\textcolor{black}{Because precise optimization is computationally too costly, we set $\gamma=1$ and set $\alpha, \beta \in \{0.4, 0.8, 1.2, 1.6\}$ to search for the optimal pair of $\alpha$ and $\beta$.}
For this parameter set, \textcolor{black}{all models yielded positive values of $\overline{\rho}(0)$, consistent with} the empirical data.
\textcolor{black}{For the GM,} $\overline{\rho}(\textcolor{black}{d})$ \textcolor{black}{decreased} \textcolor{black}{towards zero} as $\textcolor{black}{d}$ increased \textcolor{black}{for} $\textcolor{black}{d} <$ \textcolor{black}{$6$ km, i.e., the value of $\overline{\rho}(d)$ decayed faster than the empirical values}.
\textcolor{black}{At} $\textcolor{black}{d} >$ \textcolor{black}{6} km, $\overline{\rho}(\textcolor{black}{d})$ \textcolor{black}{generated by the GM was} around zero but tended to be \textcolor{black}{smaller} than the empirical \textcolor{black}{values}.
The two extended GMs yielded a decay of $\overline{\rho}(\textcolor{black}{d})$, which hit zero at $\textcolor{black}{d} \approx 20$ km\textcolor{black}{, qualitatively the same as the behaviour of} the empirical data.
The \textcolor{black}{two extended GMs} generated \textcolor{black}{larger} $\overline{\rho}(\textcolor{black}{d})$ value\textcolor{black}{s} than the empirical \textcolor{black}{values for} $\textcolor{black}{d} \le 20$ km.

To investigate the \textcolor{black}{robustness of the results against variation in the parameters of the models, we varied the parameter values as} $\alpha \in \{0.4, 0.8, 1.2, 1.6\}$ and $\beta \in \{0.4, 0.8, 1.2, 1.6\}$ \textcolor{black}{and} measured the discrepanc\textcolor{black}{y between the model and empirical data in terms of the discrepancy measure defined} by Eq.~(\ref{accuracy})\textcolor{black}{. The results for the three models are shown in} Fig.~\ref{accuracy_m}.
The data obtained from the GM was inaccurate except when $\alpha$ or $\beta$ was small. 
\textcolor{black}{In addition}, the minimum discrepancy for the GM ($=1.469$) was larger than \textcolor{black}{that} for the d-aggregate GM and s-aggregate GM ($=1.163$ and $1.161$, respectively).
The d-aggregate GM showed \textcolor{black}{a relatively} good agreement with the empirical data \textcolor{black}{in} a wide parameter region.
The performance of the s-aggregate GM was comparable with that of the d-aggregate GM \textcolor{black}{only} when $\alpha=0.4$ or $0.8$.
Our analysis suggest\textcolor{black}{s} that \textcolor{black}{aggregating nearby} cells around either of the source or destination of migration \textcolor{black}{seems to} improve the \textcolor{black}{explanatory power} of the \textcolor{black}{GM}.
The performance of the d-aggregate GM was better than that of the s-aggregate GM in terms of the robustness against variation in the parameter values.

\subsection*{\textcolor{black}{Effects of the granuity} of \textcolor{black}{spatial} aggregation}
\textcolor{black}{We set $\textcolor{black}{d}_{\rm ag}$, the \textcolor{black}{width} for \textcolor{black}{spatial smoothing} of the population density at the source or destination \textcolor{black}{cell} in the extended GM models, to 0.65 km in the previous sections.}
To investigate \textcolor{black}{the robustness of the results with respect to the $\textcolor{black}{d}_{\rm ag}$ value}, we used $\textcolor{black}{d}_{\rm ag}=1$ km, $5$ km and $25$ km \textcolor{black}{combined with the d-aggregate and s-aggregate GMs.}
The \textcolor{black}{discrepancy between each} model \textcolor{black}{and the empirical data} \textcolor{black}{is shown in Fig. 6}.

\textcolor{black}{When $\textcolor{black}{d}_{\rm ag} = 1$ km, for both} models, \textcolor{black}{the results} were similar \textcolor{black}{to those for $\textcolor{black}{d}_{\rm ag} = 0.65$ km (Fig. \ref{rho})}.
\textcolor{black}{When} $\textcolor{black}{d}_{\rm ag}=5$ and $25$ km, the behaviour of $\overline{\rho}(\textcolor{black}{d})$ was qualitatively \textcolor{black}{different, with $\overline{\rho}(\textcolor{black}{d})$ first increasing and then decreasing as $\textcolor{black}{d}$ increased, or even more complicated behaviour (i.e., s-aggregate GM with $\textcolor{black}{d}_{\rm ag}=25$ km shown in Fig. 6(b)).}

Figure~\ref{accuracy_r17} confirms \textcolor{black}{that} the \textcolor{black}{results shown in Fig.~\ref{aggregate_range}} remain \textcolor{black}{ qualitatively} the same \textcolor{black}{in} a wide range of $\alpha$ and $\beta$. 
\textcolor{black}{In other words, the results for $\textcolor{black}{d}_{\rm ag}=1$ (Figs. 7(a) and 7(b)) are similar to those for $\textcolor{black}{d}_{\rm ag}=0.65$ (Figs. 5(b) and 5(c)), whereas those for $\textcolor{black}{d}_{\rm ag}=5$ (Figs. 7(c) and 7(d)) and $\textcolor{black}{d}_{\rm ag}=25$ (Figs. 7(e) and 7(f)) are not.}
\textcolor{black}{We conclude that aggregating the population density at the source or destination of migration with $\textcolor{black}{d}_{\rm ag}$ \textcolor{black}{=} 5 km or \textcolor{black}{larger} does not even qualitatively explain the empirical data.}

\subsection*{One-dimensional toy model}
To gain further insights into the spatial inter-dependency of the population growth rate in terms of in- and out-migratory flows of populations, we analysed a toy model on the one-dimensional lattice \textcolor{black}{(i.e., chain) with} 21 cells (Fig. ~\ref{1d_scheme}).
\textcolor{black}{Differently from the simulations presented in the previous sections, the current toy model assumes a flat initial population density except in the three central cells. Combined with the simplifying assumption of the one-dimensional landscape, we aimed at revealing a minimal set of conditions under which the empirically observed patterns were produced.}
\textcolor{black}{We focussed on} the central cell and its two neighbouring cells\textcolor{black}{, one on each side on the chain}.
We set the initial number of inhabitants in the central cell to $x$, those of the two neighbouring cells to $x'$, and those of the other cells to one \textcolor{black}{as normalisation}.
The distance between two adjacent cells was \textcolor{black}{set to unity without a loss of generality.}
\textcolor{black}{Then, we} investigated the net flow \textcolor{black}{(i.e., population growth rate)}, in-flow and out-flow \textcolor{black}{of populations as a function of} $x$ and $x'$ \textcolor{black}{using the three GMs}.
We set $\textcolor{black}{d}_{\rm ag}=1$, with which we aggregated three cells \textcolor{black}{to calculate the population density at the source or destination of the immigration in the two extensions of the GM}.

The net flow, in-flow and out-flow \textcolor{black}{in} the \textcolor{black}{three models} are shown in Fig.~9.
In the GM, the \textcolor{black}{net flow at the central cell} heavily depended on $x$ but negatively and \textcolor{black}{only} slightly depend\textcolor{black}{ed} on \textcolor{black}{the} population size in the neighbouring cells $x'$ (Fig.~\ref{1d_model}(a)). 
This result was inconsistent with the empirical\textcolor{black}{ly observed} pattern (Fig.~\ref{rho}). 
This \textcolor{black}{inconsistency} was due to an increase in the out-flow \textcolor{black}{at the central cell as} $x'$ \textcolor{black}{increased} (Fig.~\ref{1d_model}(c)), whereas the in-flow \textcolor{black}{at the central cell was not sensitive to} $x'$ (Fig.~9(b)).

The patterns of migration flows for the d-aggregate and s-aggregate GMs were qualitatively different from those for the GM (Figs.~\ref{1d_model}(d)--(i)). 
\textcolor{black}{In both models, the population growth rate increased as} $x'$ \textcolor{black}{increased (Figs.~9(d) and 9(g)), which is consistent with the empirically observed patterns.}
In the d-aggregate GM, this change \textcolor{black}{mainly owed to} changes in the in-flow\textcolor{black}{, which increased as $x'$ increased (Fig. 9(e))}. 
The out-flow for the d-aggregate GM was similar to that for the GM (Fig.~\ref{1d_model}(f)).
\textcolor{black}{In other words,} a cell \textcolor{black}{surrounded by those with higher population density} attract\textcolor{black}{ed a larger} migration flow \textcolor{black}{in the d-aggregate GM.}
\textcolor{black}{In contrast, in the s-aggregate GM,} change\textcolor{black}{s in the population flow were} mainly attribut\textcolor{black}{ed} to changes in the out-flow.
The in-flow for the d-aggregate GM was similar to that for the GM (Fig.~\ref{1d_model}(h)) \textcolor{black}{and} the out-flow decreased \textcolor{black}{as} $x'$ \textcolor{black}{increased} for the d-aggregate GM (Fig.~\ref{1d_model}(i)).
\textcolor{black}{In other words,} a cell surrounded by \textcolor{black}{those with higher population density} was less likely to \textcolor{black}{lose inhabitants in the s-aggregate GM}.

\subsection*{GM with the aggregation around both the source and destination \textcolor{black}{cells}}
Lastly, we investigated \textcolor{black}{an extension of the} GM with the aggregation \textcolor{black}{of cells} around both the source and destination cells\textcolor{black}{, called the sd-aggregate GM (Appendix E)}.
The behaviour of $\overline{\rho}(\textcolor{black}{d})$ was \textcolor{black}{quaitatively the same as} that obtained from the d-aggregate GM, s-aggregate GM \textcolor{black}{and empirical data} (Fig.~18).
In addition, \textcolor{black}{the sd-aggregate GM} was accurate \textcolor{black}{in} a wide parameter \textcolor{black}{region} (Fig.~19).
We also confirmed that the discrepanc\textcolor{black}{y measure for the sd-aggregate GM} increased as $\textcolor{black}{d}_{\rm ag}$ increased (Figs.~\ref{aggregate_range_sd-aggregate} and \ref{accuracy_range_sd-aggregate})\textcolor{black}{, similar to the results for the d-aggregate and s-aggregate GMs (Figs.~6 and 7).}
The \textcolor{black}{behaviour of this model on the one-dimensional toy model} was also consistent with the empirical data \textcolor{black}{(Fig.~22)} because the in-flow and out-flow of the model were similar to those for the d-aggregate GM and s-aggregate GM, respectively.

\section*{Discussion}
\textcolor{black}{W}e investigated spatial patterns of \textcolor{black}{demographic dynamics through the analysis of} the population census \textcolor{black}{data in} Japan \textcolor{black}{in 2005 and 2010}.
We found that the \textcolor{black}{population} growth rate \textcolor{black}{in a cell} was positively correlated with the population density in \textcolor{black}{cells nearby, in addition to that in the focal cell.}
We used the gravity model and its \textcolor{black}{variants} to investigate \textcolor{black}{possible} effects of migration on \textcolor{black}{the empirically observed} spatial patterns of \textcolor{black}{the} population \textcolor{black}{growth rate}.
Under the framework of the GM, we found that aggregating some neighbouring cells around either the source or destination \textcolor{black}{of} migration \textcolor{black}{events considerably} improve\textcolor{black}{d} the \textcolor{black}{fit} of the \textcolor{black}{GM model to the empirical data}.
\textcolor{black}{The results were better when the cells around} the destination cell \textcolor{black}{were aggregated, in particular regarding the robustness of the results against variation in the parameter values,} \textcolor{black}{than when the cells around the} source cell \textcolor{black}{were aggregated}.
\textcolor{black}{All the results were qualitatively the same when we set $t_1=2000$ and $t_2 = 2005$, although the census data in 2000 were less accurate than those in 2005 and 2010} \textcolor{black}{(Appendix A). }

\textcolor{black}{Aggregation of cells near the destination cell models behaviour of individual\textcolor{black}{s that} perceive the population of the destination cell as a sum (or average) of the population over the cells neighbouring the destination cell. Because the size of the cell is imposed by the empirical data, aggregation of cells around the destination \textcolor{black}{cell} is equivalent to decreasing the spatial resolution of the GM by coarse graining}.
\textcolor{black}{Traditionally, administrative boundaries have been used as operational units of the GM \cite{Barthelemy2011}.
 A cluster identified by the city clustering algorithm may also be used as the unit \cite{Rozenfeld2008, Rozenfeld2009}. In the continuous-space GM, the unit is assumed to be an infinitely small spatial segment \cite{Simini2013}. However, there is no a priori reason to assume that any one of these units is an appropriate choice.}
\textcolor{black}{Our results suggest that \textcolor{black}{spatial averaging} with \textcolor{black}{a circle of} radius $\textcolor{black}{d}_{\rm ag} \approx 1$ km may be a reasonable choice as compared to a larger $\textcolor{black}{d}_{\rm ag}$ or the original cell size (i.e., $500 \times 500$ $\rm m^2$). Real inhabitants may perceive the population density at the destination as a spatial average on this scale. Although we reached this conclusion using the GMs, this guideline may be also useful when other migration models are used.}

The present study has limitations.
First, due to \textcolor{black}{a high} computational \textcolor{black}{cost}, we only examined a limited number of combinations of parameter \textcolor{black}{values in} the \textcolor{black}{GMs}.
\textcolor{black}{A more exhaustive search of the parameter space or the use of different migration models, as well as analyzing different data sets, warrants future work.}

Second, due to the lack of empirical data, we \textcolor{black}{could not analyse} more microscopic processes \textcolor{black}{contributing to} population changes.
For example, because of the absence of \textcolor{black}{spatially explicit} data on \textcolor{black}{the number of births and deaths}, we \textcolor{black}{did not include} births and deaths into \textcolor{black}{our} models.
However, the observed in-flow and out-flow were at least twice as large as the numbers of births and deaths in all the 47 prefectures in Japan (Table 2). 
Therefore, migration rather than births and deaths seems to be a main driver of spatially untangled population changes in Japan during the observation period.
The lack of data also prohibited us from looking into the effect of the age \textcolor{black}{of inhabitants. In fact,} individuals at a certain life stage are more likely to migrate in general \cite{Lewis1982, Kahley1991}. Data on migration flows between cells, \textcolor{black}{births, deaths and the age distribution,} which \textcolor{black}{are} not included in \textcolor{black}{the present} data set, \textcolor{black}{are expected to enable further investigations of the spatial patterns of population changes examined in the present study}.

\textcolor{black}{Third, our conclusions are based on the longitudinal data at only two time points in a single country. The strength of the current results should be understood as such.}

\textcolor{black}{Fourth, we did not take into account the effect of water-surface cells, which cannot be inhabited. The population density at distance $d$ from a focal cell $i$, i.e., $D_i(d)$, is therefore underestimated when cell $i$ is located near water (e.g., sea, lake, large river). Additional information about the geographical property of cells such as the water area within the cell and the land use may improve the present analysis.}

\newpage
\subsection*{Ethics}
This study required no ethical approval because all relevant data are
publicly available.

\subsection*{Data availability}
Data are available from \url{http://e-stat.go.jp/SG2/eStatGIS/page/download.html}.

\subsection*{Authors' contributions}
NM designed the study. KT collected and analysed the data. 
All authors wrote the manuscript.

\subsection*{Competing interests}
The authors declare that they have no competing interests.

\subsection*{Funding}
This work is supported by JST, CREST (No. JPMJCR1304). 
NM acknowledges the support provided through JST, ERATO, Kawarabayashi Large Graph Project.

\newpage
\section*{Appendix A: Population changes between 2000 and 2005}
\def\theequation{A\arabic{equation}}
\renewcommand{\figurename}{Figure}
\makeatletter
\renewcommand{\thefigure}{A\arabic{figure}}
\renewcommand{\thetable}{A\arabic{table}}
\makeatother
\setcounter{table}{0}
\setcounter{figure}{0}
\setcounter{equation}{0}
In the main text, we used \textcolor{black}{the} data on the \textcolor{black}{p}opulation \textcolor{black}{c}ensus in Japan in 2005 and 2010.
\textcolor{black}{The data} on 2000 are also \textcolor{black}{publicly} available \textcolor{black}{although they} \textcolor{black}{are less} accurate \textcolor{black}{than} those in 2005 and 2010 \cite{gaiyo}. 
\textcolor{black}{Here we set $t_1 = 2000$ and $t_2 = 2005$ and ran} the same analysis \textcolor{black}{pipeline} with that in the main text \textcolor{black}{to examine the robustness of our results.}
\textcolor{black}{As shown in the following, the} results were qualitatively the same as those shown in the main text for ($t_1$, $t_2$) $=$ (2005, 2010) (Figs.~4--7)\textcolor{black}{, except for the behaviour of the GM}.

In Fig.~\ref{rho12}, $\overline{\rho}(\textcolor{black}{d})$ obtained from the empirical \textcolor{black}{data,} the GM, d-aggregate GM and s-aggregate GM is compared.
\textcolor{black}{Similarly to the analysis shown in the main text, for the three GMs, we set $\gamma=1$ and varied $\alpha$, $\beta\in \{ 0.4, 0.8, 1.2, 1.6\}$ and used the optimized parameter values.}
The $\overline{\rho}(0)$ value for the GM was \textcolor{black}{negative, contradicting} the empirical data, whereas the behaviour of the d-aggregate and s-aggregate GMs \textcolor{black}{was} qualitatively the same as that of the empirical data.

For $\alpha \in \{0.4, 0.8, 1.2, 1.6\}$ and $\beta \in \{0.4, 0.8, 1.2, 1.6\}$\textcolor{black}{, the} discrepanc\textcolor{black}{y} between the model and empirical data (Eq.~(\ref{accuracy})) is shown in Fig.~\ref{accuracy_m12}.
The \textcolor{black}{results for} the GM was inaccurate for all parameter combinations that we considered (Fig.~\ref{accuracy_m12}(a)).
The d-aggregate GM yielded a good agreement with the data in a wide parameter region (Fig.~\ref{accuracy_m12}(b))\textcolor{black}{. The} s-aggregate GM was accurate only for $\alpha=0.4$ (Fig.~\ref{accuracy_m12}(c)).
These results are similar to those for ($t_1$, $t_2$) = (2005, 2010) (Fig.~5).

We \textcolor{black}{then} examined the robustness of the results with respect to the $\textcolor{black}{d}_{\rm ag}$ value.
The discrepanc\textcolor{black}{y} between the models and the empirical data is shown in Fig.~\ref{rho_aggregate_range12}.
For both \textcolor{black}{d-aggregate and s-aggregate GMs}, $\overline{\rho}(\textcolor{black}{d})$ \textcolor{black}{behaved similarly to that for the empirical data when $\textcolor{black}{d}_{\rm ag} =1$ km but not when $\textcolor{black}{d}_{\rm ag}=5$ km and 25 km}.
Figure~\ref{accuracy_range12} confirms this result for various values of $\alpha$ and $\beta$.
For a wide region of the $\alpha$-$\beta$ parameter space, the discrepancy increased as $\textcolor{black}{d}_{\rm ag}$ increased.

\section*{Appendix B: \textcolor{black}{Plot of} $\rho_k(\textcolor{black}{d})$ for \textcolor{black}{each 50 km $\times$ 50 km} region}
\def\theequation{B\arabic{equation}}
\renewcommand{\figurename}{Figure}
\makeatletter
\renewcommand{\thefigure}{B\arabic{figure}}
\renewcommand{\thetable}{B\arabic{table}}
\makeatother
\setcounter{table}{0}
\setcounter{figure}{0}
\setcounter{equation}{0}

In the main text, we showed the values of $\rho_{k}(\textcolor{black}{d})$ averaged over \textcolor{black}{all} regions of size 50 km $\times$ 50 km, denoted by, $\overline{\rho}(\textcolor{black}{d})$ (Fig.~4).
\textcolor{black}{The} $\rho_{k}(\textcolor{black}{d})$ for each region $k$ \textcolor{black}{is plotted as a function of $\textcolor{black}{d}$ in Fig.~14.}
\textcolor{black}{The} values of $\rho_{k}(\textcolor{black}{d})$ \textcolor{black}{considerably depend on the region.}

\textcolor{black}{To calculate $\overline{\rho}(d)$, we used all regions. However, some regions and their nearby regions include water-surface cells, potentially biasing the estimation of $\overline{\rho}(d)$. Therefore, we examined the $\rho_{k}(d)$ values for region $k$ such that all cells within region $k$ and those within 30 km from any cell in region $k$ are not in the sea.}
\textcolor{black}{The average of $\rho_k(d)$ over these regions is qualitatively the same as that shown in the main text (Fig.~15).}

\section*{Appendix C: \textcolor{black}{Analysis of} cells \textcolor{black}{with more than 100 inhabitants}}
\def\theequation{C\arabic{equation}}
\renewcommand{\figurename}{Figure}
\makeatletter
\renewcommand{\thefigure}{C\arabic{figure}}
\renewcommand{\thetable}{C\arabic{table}}
\makeatother
\setcounter{table}{0}
\setcounter{figure}{0}
\setcounter{equation}{0}

In the main text, we used cells whose population size was between 10 and 100. 
Figure~\ref{rho_over100} shows $\overline{\rho}(\textcolor{black}{d})$ for cells whose population size was greater than 100.
The behaviour of $\overline{\rho}(\textcolor{black}{d})$ was qualitatively the same as that for the cells of the population size between 10 and 100 (Fig.~\ref{rho}).

\section*{Appendix D: \textcolor{black}{Analysis with} the partial correlation \textcolor{black}{coefficient}}
\def\theequation{D\arabic{equation}}
\renewcommand{\figurename}{Figure}
\makeatletter
\renewcommand{\thefigure}{D\arabic{figure}}
\renewcommand{\thetable}{D\arabic{table}}
\makeatother
\setcounter{table}{0}
\setcounter{figure}{0}
\setcounter{equation}{0}

In the main text, we calculated $\rho_{k}(\textcolor{black}{d})$ using the \textcolor{black}{Pearson} correlation coefficient (Eq.~\eqref{rho}). However, the strong spatial correlation in the population size \textcolor{black}{combined with the tendency that a highly populated cell grows more than sparsely populated cells do may} result in spurious\textcolor{black}{ly large $\rho_k(\textcolor{black}{d})$ values}. Therefore, to control for the spatial correlation in the population size, we calculated the partial correlation coefficient between the \textcolor{black}{population} growth rate \textcolor{black}{of a cell} and the population density in nearby cells, $\rho_{k}^{\prime}(\textcolor{black}{d})$, by 
\begin{equation}
\rho_{k}^{\prime}(\textcolor{black}{d}) = \frac{ \rho_{k}(\textcolor{black}{d}) - \text{ cor}_k(D_{i}(\textcolor{black}{d}), n_{i}) \times \text{ cor}_k(R_{i}, n_{i})}{\sqrt{1- \text{ cor}_k(D_{i}(\textcolor{black}{d}),n_{i})} \sqrt{1- \text{ cor}_k(R_{i},n_{i})}},
\label{eq:partial_rho}
\end{equation}
where $\text{ cor}_k(\cdot, \cdot)$ is the Pearson correlation coefficient between two variables in region $k$; and $D_{i}(\textcolor{black}{d})$ is the population density averaged over the cells at distance $\textcolor{black}{d}$ from cell $i$ in region $k$; $R_i$ is the population growth rate of cell $i$ in region $k$; $n_i$ is the number of inhabitants in cell $i$ in region $k$. 
We defined $\overline{\rho}^{\prime}(\textcolor{black}{d})$ as the average of $\rho_{k}^{\prime}(\textcolor{black}{d})$ over all regions. 
\textcolor{black}{Figure}~\ref{partial_cor} shows that $\overline{\rho}^{\prime}(\textcolor{black}{d})$ as a function of $\textcolor{black}{d}$ \textcolor{black}{ behaves similarly to $\overline{\rho}(\textcolor{black}{d})$ does }(Fig.~\ref{rho}).

\section*{Appendix E: GM with the population density \textcolor{black}{aggregated around both} the source and destination \textcolor{black}{cells}}
\def\theequation{E\arabic{equation}}
\renewcommand{\figurename}{Figure}
\makeatletter
\renewcommand{\thefigure}{E\arabic{figure}}
\renewcommand{\thetable}{E\arabic{table}}
\makeatother
\setcounter{table}{0}
\setcounter{figure}{0}
\setcounter{equation}{0}
In the main text, we aggregated \textcolor{black}{the} cells around either the source or destination \textcolor{black}{cell but not both.}
Here we carried out aggregation around both the source and destination \textcolor{black}{cells}.
In the model, which we refer to as the GM with the aggregation around both the source and destination (sd-aggregate GM), the population flow from cell $i$ to cell $j$ is \textcolor{black}{defined} by
\begin{equation}\label{eq_sd-aggregate}
T_{ij} = G n_{i} \frac{N_{i}(\textcolor{black}{d}_{\rm ag})^{\alpha-1} N_{j}(\textcolor{black}{d}_{\rm ag})^{\beta}}{d_{ij}^{\gamma}}.
\end{equation}
\textcolor{black}{As we present in the following, the} behaviour of the sd-aggregate GM was similar to that of the d-aggregate GM (Figs.~\ref{rho}--\ref{accuracy_r17}).

\textcolor{black}{We compare} $\overline{\rho}(\textcolor{black}{d})$ \textcolor{black}{between} the empirical and \textcolor{black}{simulated} data \textcolor{black}{in Fig.~18}.
\textcolor{black}{The behaviour of} $\overline{\rho}(\textcolor{black}{d})$ \textcolor{black}{obtained} from the GM w\textcolor{black}{as} qualitatively the same as \textcolor{black}{that} of the empirical data.
\textcolor{black}{The discrepancy between the model and empirical data (Eq.~(8))} w\textcolor{black}{as} \textcolor{black}{small in} a wide parameter \textcolor{black}{region} (Fig.~\ref{accuracy_sd-aggregate}).
We \textcolor{black}{also} confirmed that the discrepanc\textcolor{black}{y} increased as $\textcolor{black}{d}_{\rm ag}$ increased (Figs.~\ref{aggregate_range_sd-aggregate} and \ref{accuracy_range_sd-aggregate}).
The net flow, in-flow and out-flow \textcolor{black}{for} the sd-aggregate GM \textcolor{black}{simulated} on a chain with 21 cells are shown in Fig.~22.
The in-flow and out-flow for the sd-aggregate GM \textcolor{black}{(Figs.~22(b) and 22(c))} were similar to those for the d-aggregate GM (Fig. 9(e)) and the s-aggregate GM (Fig. 9(i)), respectively.
As a result, the net-flow \textcolor{black}{for} the sd-aggregate GM (Fig.~\ref{flow_sd-aggregate}(a)) was similar to \textcolor{black}{that for} the d-aggregate GM (Fig.~\ref{1d_model}(d)) and the s-aggregate GM (Fig.~\ref{1d_model}(g)).

\newpage
\section*{Figures}
\def\theequation{}
\renewcommand{\figurename}{Figure}
\makeatletter
\renewcommand{\thefigure}{\arabic{figure}}
\renewcommand{\thetable}{\arabic{table}}
\makeatother
\setcounter{table}{0}
\setcounter{figure}{0}
\setcounter{equation}{0}

\begin{figure}[H]
\begin{center}
\includegraphics[width=140mm]{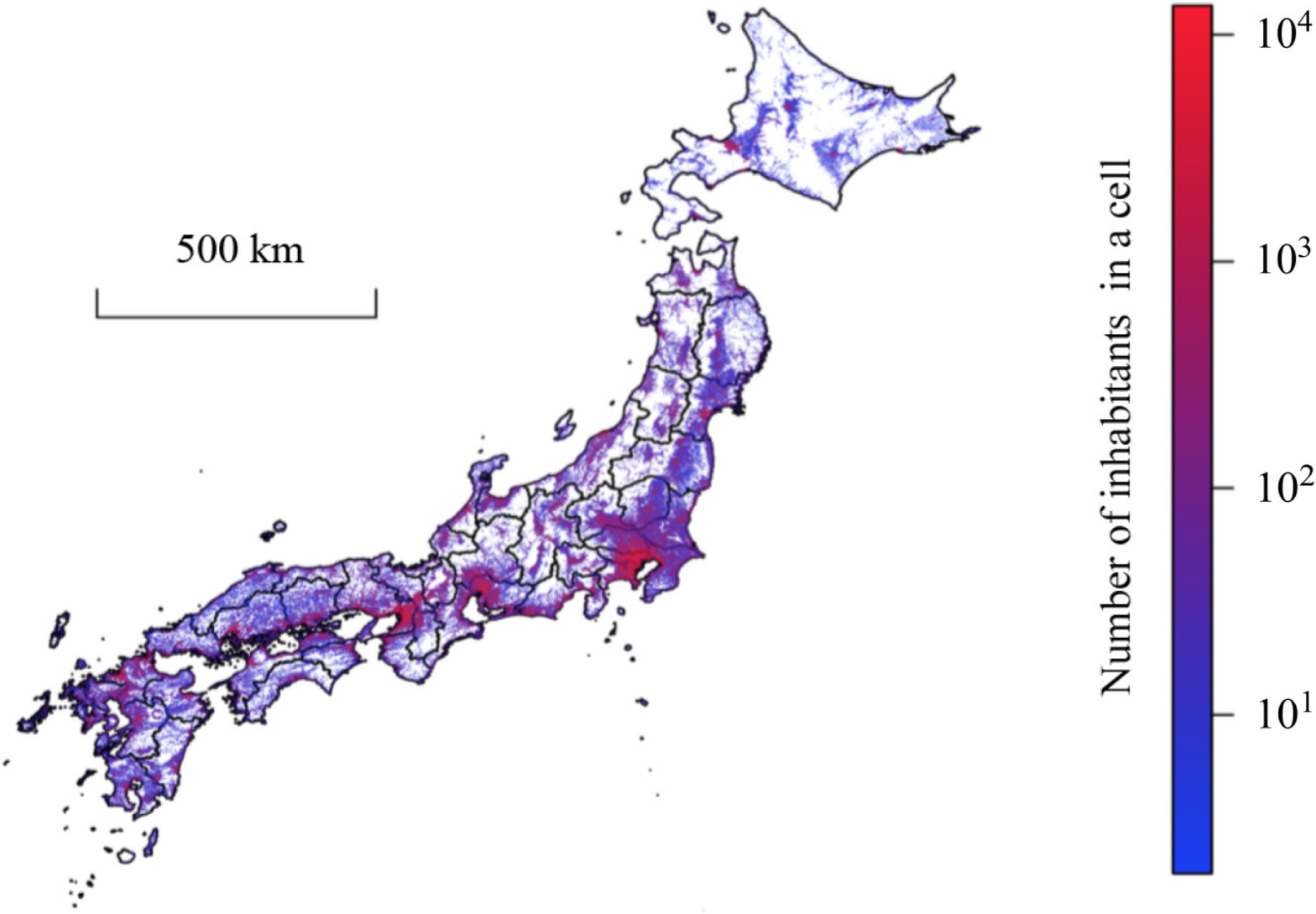}
\end{center}
\caption{The distribution of inhabitants at time $t_2$ \textcolor{black}{(i.e., year 2010)}. \textcolor{black}{The colour code} represen\textcolor{black}{ts the }numbers of inhabitants \textcolor{black}{in a cell}. \textcolor{black}{Vacant cells} are \textcolor{black}{shown in} white.}
\label{map}
\end{figure}

\newpage
\begin{figure}[H]
\begin{center}
\includegraphics[width=140mm]{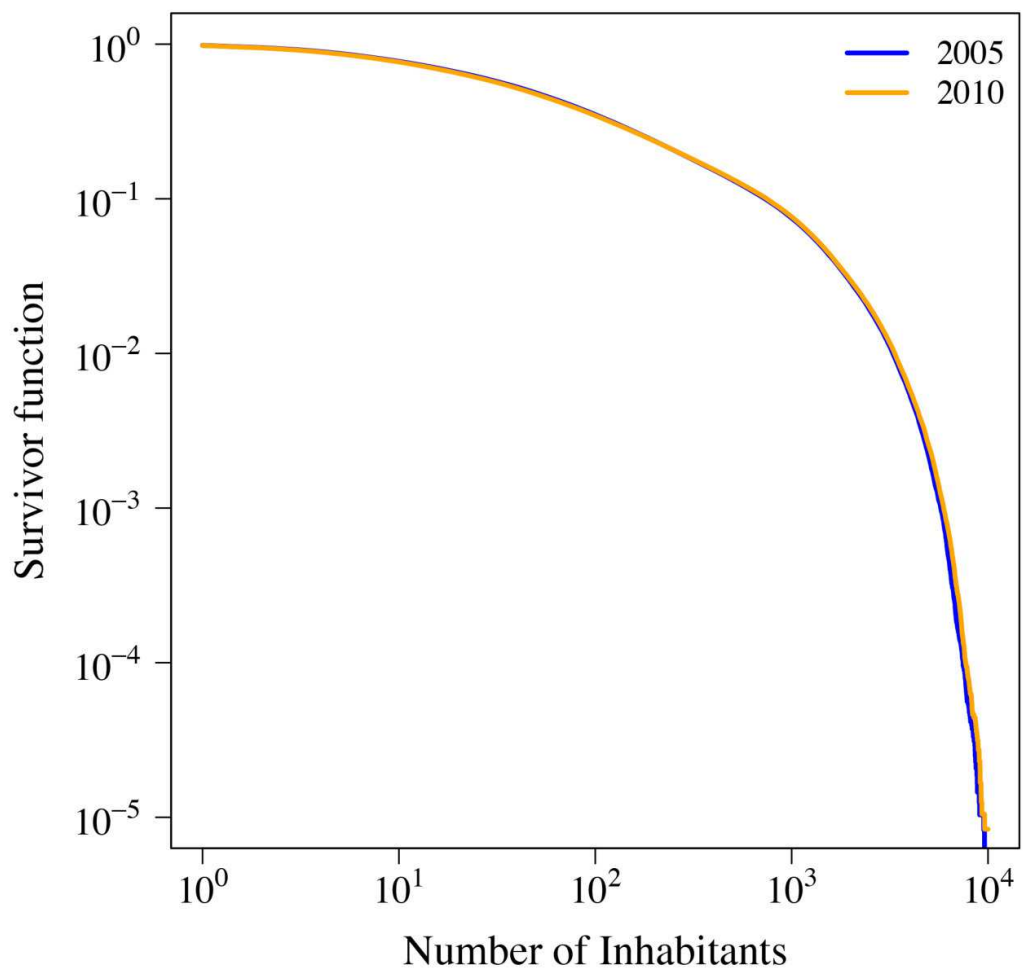}
\end{center}
\caption{The survivor function of the number of inhabitants in a cell. \textcolor{black}{The two lines almost overlap with each other.}}
\label{pop_dist}
\end{figure}

\newpage
\begin{figure}[H]
\begin{center}
\includegraphics[width=140mm]{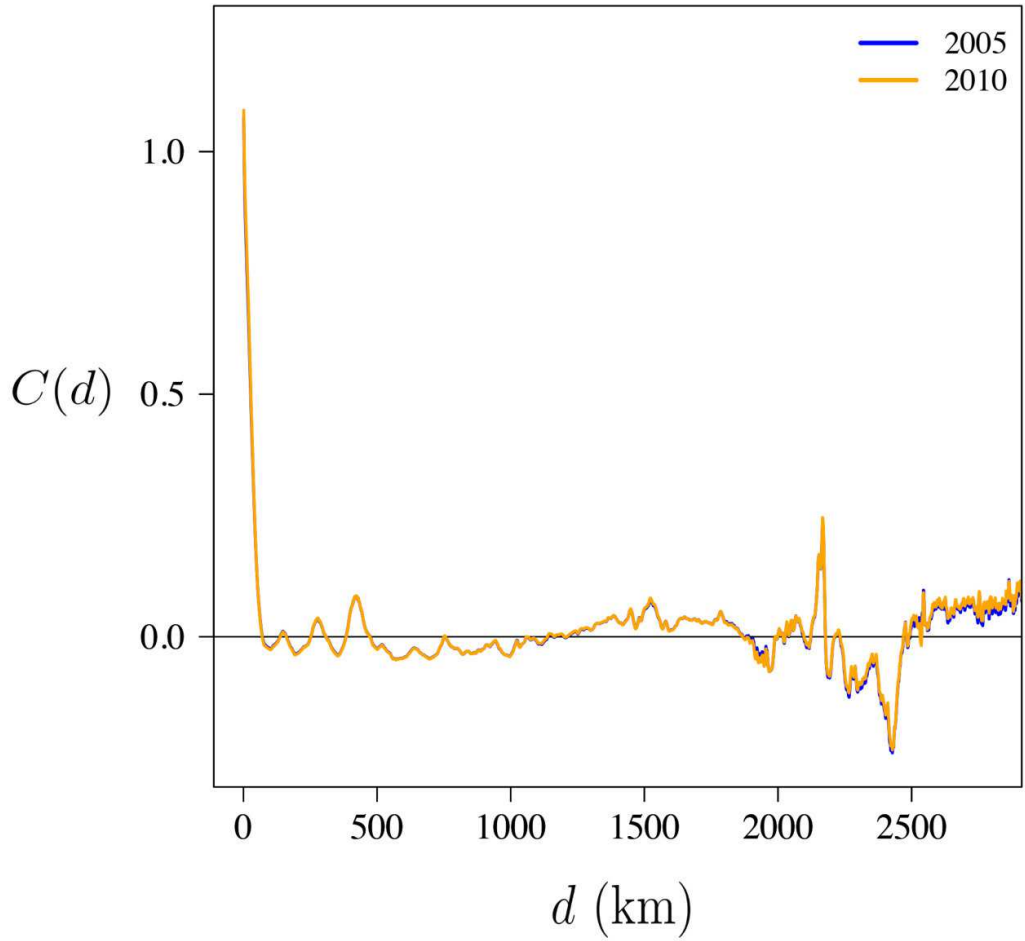}
\end{center}
\caption{The spatial correlation in \textcolor{black}{the} \textcolor{black}{number of inhabitants in the} cell. \textcolor{black}{The correlation measure $C(\textcolor{black}{d})$ is defined by Eq.~(1), and $\textcolor{black}{d}$ is the distance between the two cells. The two lines almost overlap with each other.}}
\label{spatial_cor}
\end{figure}

\newpage
\begin{figure}[H]
\begin{center}
\includegraphics[width=150mm]{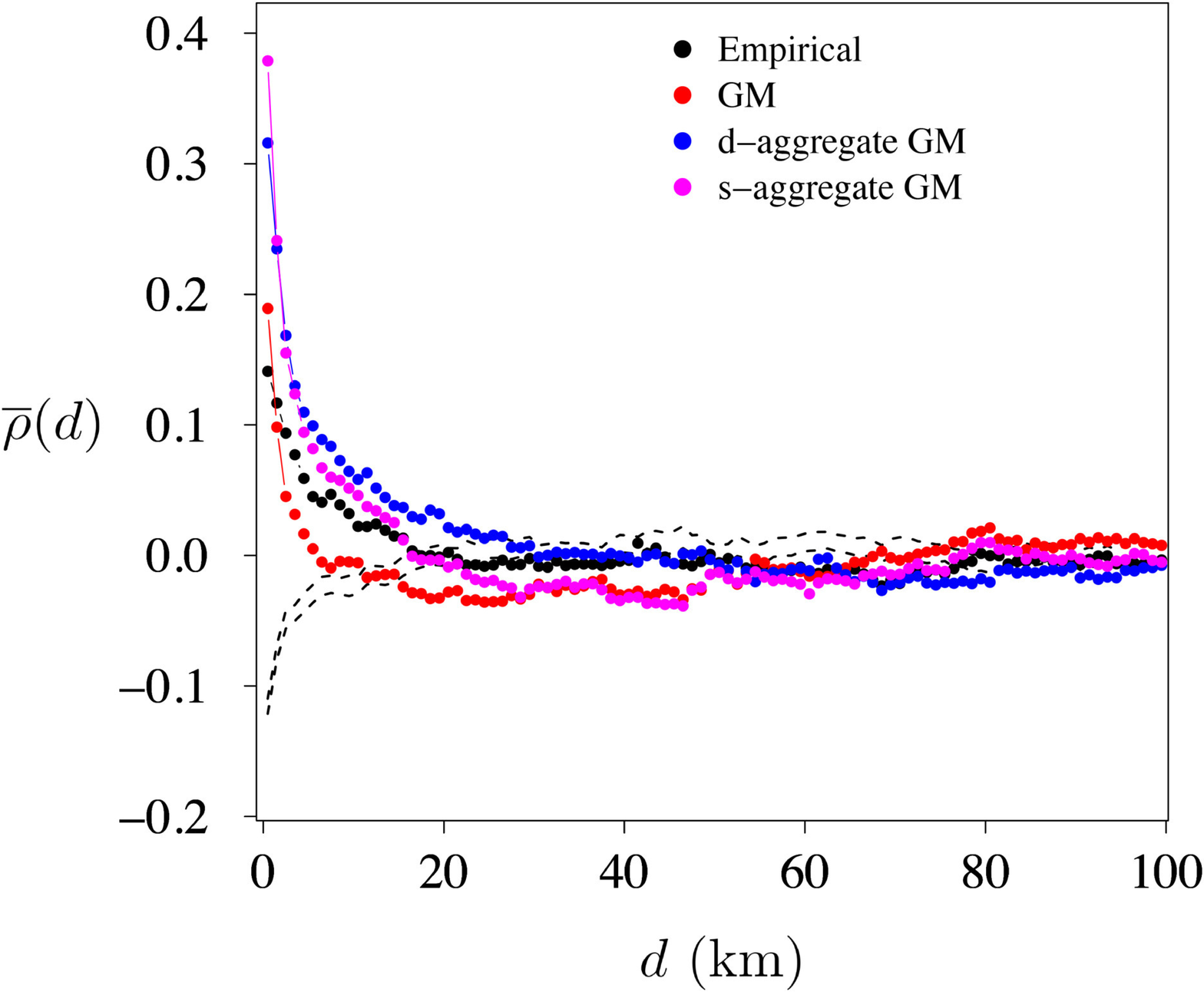}
\end{center}
\caption{Dependence of the population growth rate in a cell on the population density at distance $\textcolor{black}{d}$, $\overline{\rho}(\textcolor{black}{d})$. We set $\alpha=$ \textcolor{black}{0.4}, $\beta=$ \textcolor{black}{0.8} \textcolor{black}{and $\gamma=1$ for the GM; $\alpha = 0.8$\textcolor{black}{,} $\beta = 0.4$, $\gamma=1$ and $d_{ag} = 0.65$ km for the d-aggregate GM; $\alpha = 0.4$\textcolor{black}{,} $\beta = 0.4$, $\gamma=1$ and $d_{ag}=0.65$ km for the s-aggregate GM.}  \textcolor{black}{The ranges indicated by} the dashed lines represent 95\% \textcolor{black}{confidence intervals (CIs) generated by spatially random distributions of the number of inhabitants on the inhabited cells}.}
\label{rho}
\end{figure}

\newpage
\begin{figure}[H]
\begin{center}
\includegraphics[width=160mm]{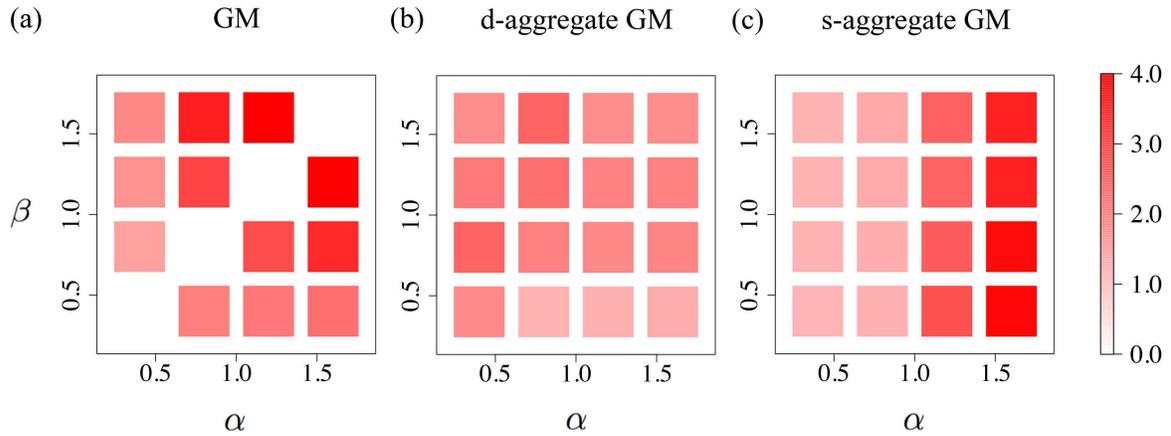}
\end{center}
\caption{The discrepancy of the GM, d-aggregate GM and s-aggregate GM \textcolor{black}{from the empirical data in terms of the discrepancy measure given by Eq.~(8)}. \textcolor{black}{A dark hue represents a large discrepancy value}. (a) GM. (b) d-aggregate GM. (c) s-aggregate GM. \textcolor{black}{The d}iagonals in (a) are blank because the in-flow and out-flow are \textcolor{black}{the same} when $\alpha = \beta$ in the GM\textcolor{black}{, resulting in a zero population growth rate in all cells}. We set $\gamma = 1$ \textcolor{black}{and $\textcolor{black}{d}_{ag}=0.65$ km}.}
\label{accuracy_m}
\end{figure}

\newpage
\begin{figure}[H]
\begin{center}
\includegraphics[width=160mm]{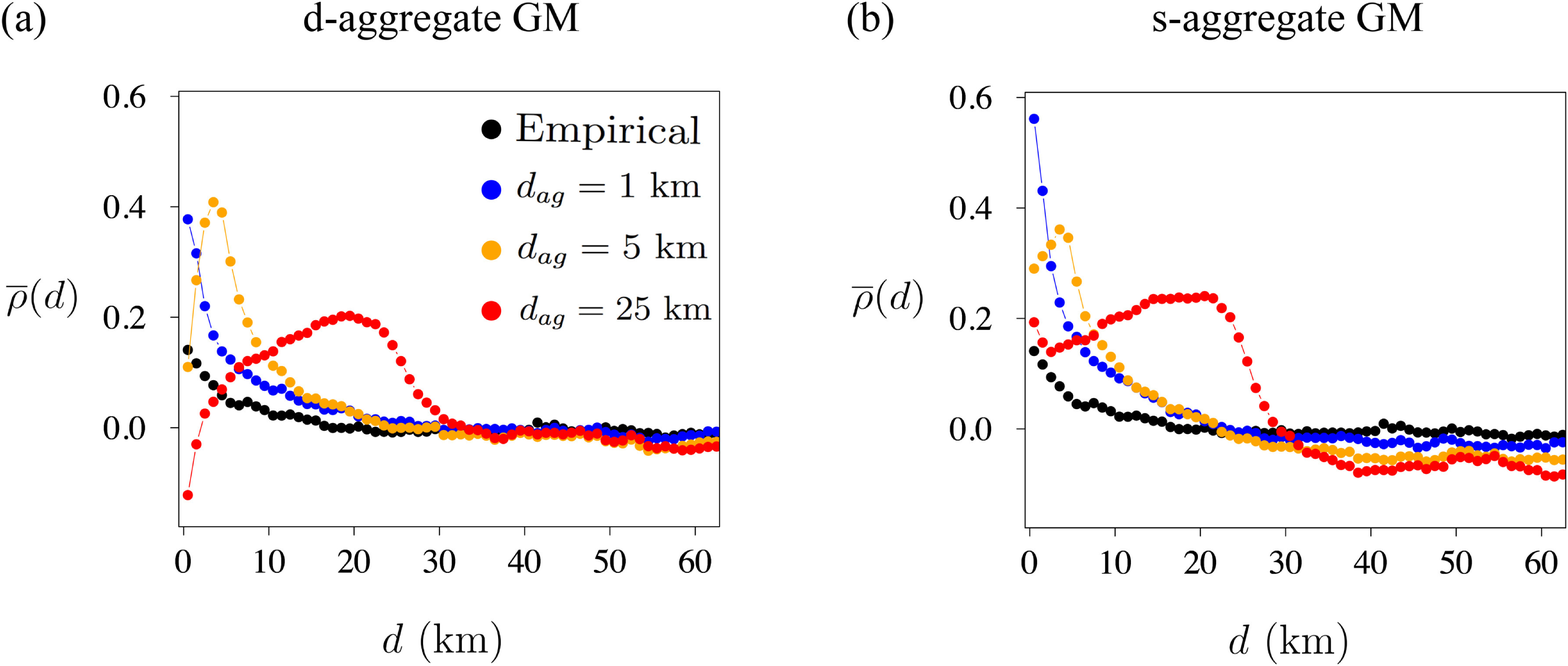}
\end{center}
\caption{Dependence of the \textcolor{black}{population} growth rate in a cell on the population density at distance $\textcolor{black}{d}$, $\overline{\rho}(\textcolor{black}{d})$, calculated from the empirical and numerical data for different values of $\textcolor{black}{d}_{\rm ag}$. (a) d-aggregate GM. \textcolor{black}{We set $\alpha=0.8$, $\beta=0.4$ and $\gamma = 1.0$.} (b) s-aggregate GM. \textcolor{black}{We set $\alpha=0.4$, $\beta=0.4$ and $\gamma = 1.0$.}}
\label{aggregate_range}
\end{figure}

\newpage
\begin{figure}[H]
\begin{center}
\includegraphics[width=130mm]{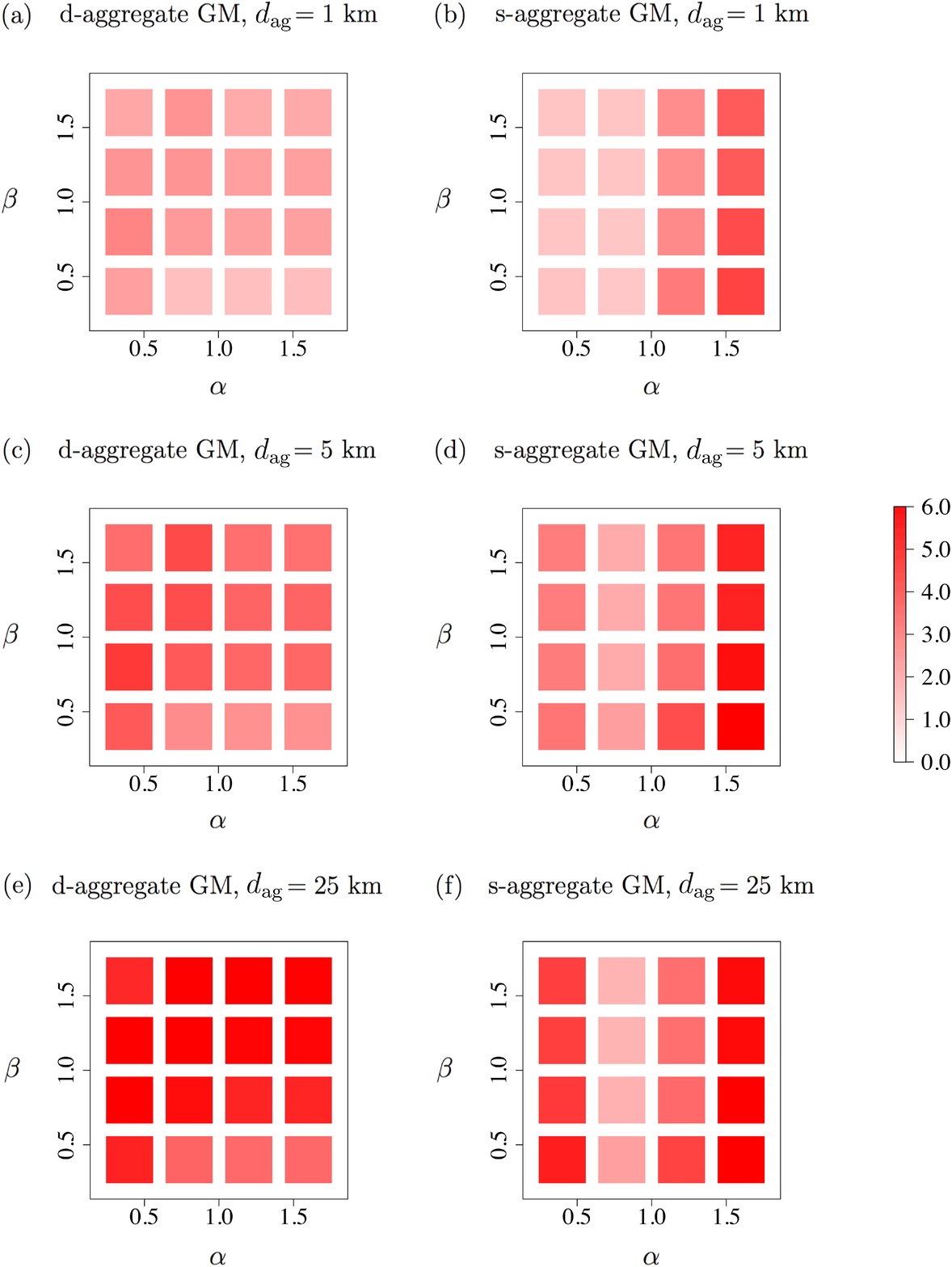}
\end{center}
\caption{The discrepancy of the d-aggregate GM and s-aggregate GM from the empirical data. (a) d-aggregate GM, $\textcolor{black}{d}_{\rm ag}=1$ km. (b) s-aggregate GM, $\textcolor{black}{d}_{\rm ag}=1$ km. (c) d-aggregate GM, $\textcolor{black}{d}_{\rm ag}=5$ km. (d) s-aggregate GM, $\textcolor{black}{d}_{\rm ag}=5$ km. (e) d-aggregate GM, $\textcolor{black}{d}_{\rm ag}=25$ km. (f) s-aggregate GM, $\textcolor{black}{d}_{\rm ag}=25$ km. }
\label{accuracy_r17}
\end{figure}

\newpage
\begin{figure}[H]
\begin{center}
\includegraphics[width=140mm]{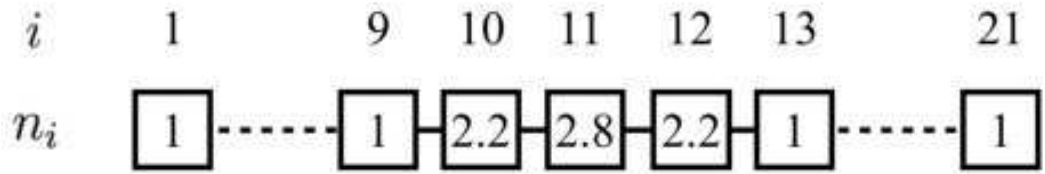}
\end{center}
\caption{The schematic of \textcolor{black}{the GM models on a} chain. A square represents a cell\textcolor{black}{, and} $n_{i}$ is the \textcolor{black}{initial} number of inhabitants in cell $i$. \textcolor{black}{We} set $x=2.8$ and $x^{\prime}=2.2$ \textcolor{black}{for illustration}. }
\label{1d_scheme}
\end{figure}

\newpage
\begin{figure}[H]
\begin{center}
\includegraphics[width=140mm]{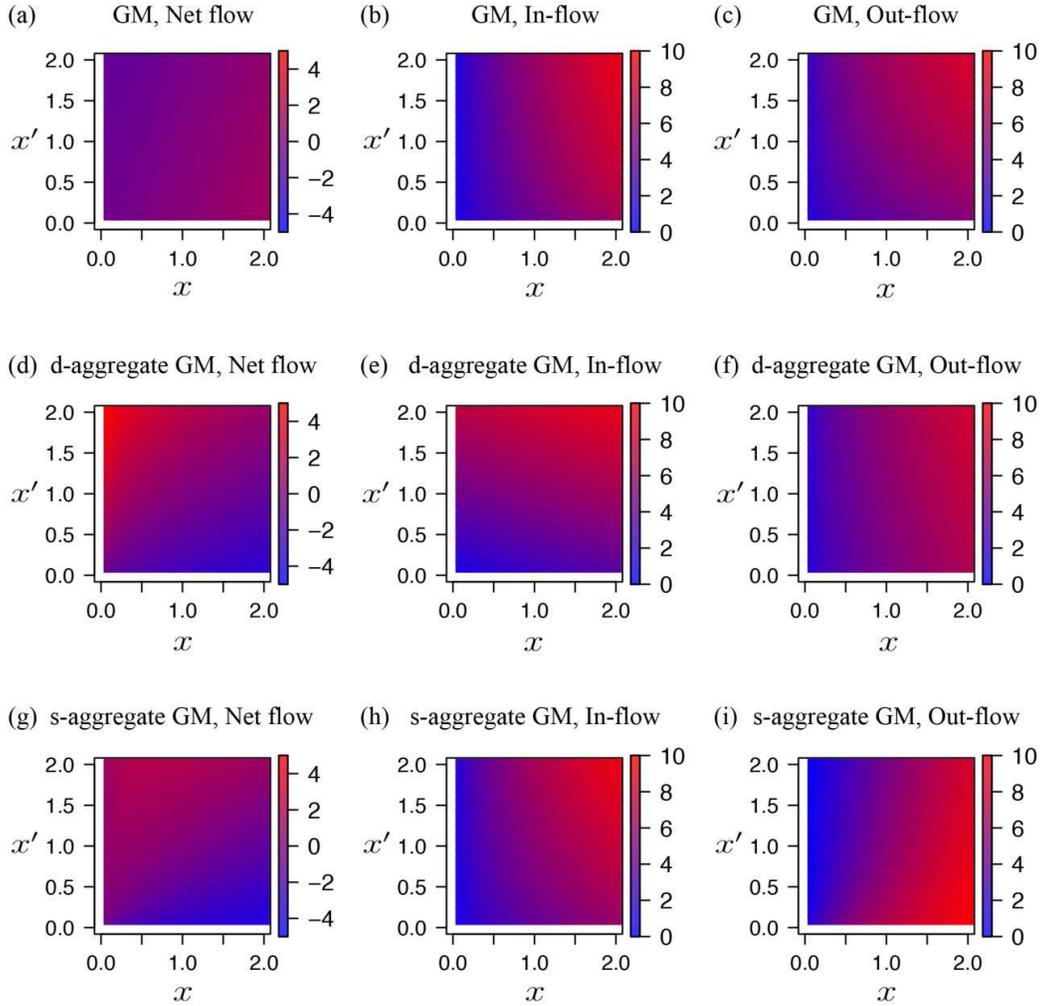}
\end{center}
\caption{The net flow, in-flow and out-flow for the GM, d-aggregate GM \textcolor{black}{and s-aggregate GM in the one-dimensional model with 21 cells. The initial condition is a symmetric distribution of the density of inhabitants that is uniform except in the central three cells. The initial population density is equal to $x$ in the central cell, $x'$ in the neighbouring two cells and 1 in the other cells. We set} $G$ for the GM, d-aggregate GM and s-aggregate GM \textcolor{black}{to} $1$, $(1/3)^{\beta}$ and $(1/3)^{\alpha-1}$, respectively, and $\alpha=0.4$, $\beta = 0.6$ and $\gamma = 1.0$. (a) Net flow \textcolor{black}{for} the GM. (b) In-flow \textcolor{black}{for} the GM. (c) Out-flow \textcolor{black}{for} the GM. (d) Net flow \textcolor{black}{for} the d-aggregate GM. (e) In-flow \textcolor{black}{for} the d-aggregate GM. (f) Out-flow \textcolor{black}{for} the d-aggregate GM. (g) Net flow \textcolor{black}{for} the s-aggregate GM. (h) In-flow \textcolor{black}{for} the s-aggregate GM. (i) Out-flow \textcolor{black}{for} the s-aggregate GM. }
\label{1d_model}
\end{figure}

\def\theequation{A \arabic{equation}}
\makeatother
\setcounter{equation}{0}

\newpage
\begin{figure}[H]
\begin{center}
\includegraphics[width=150mm]{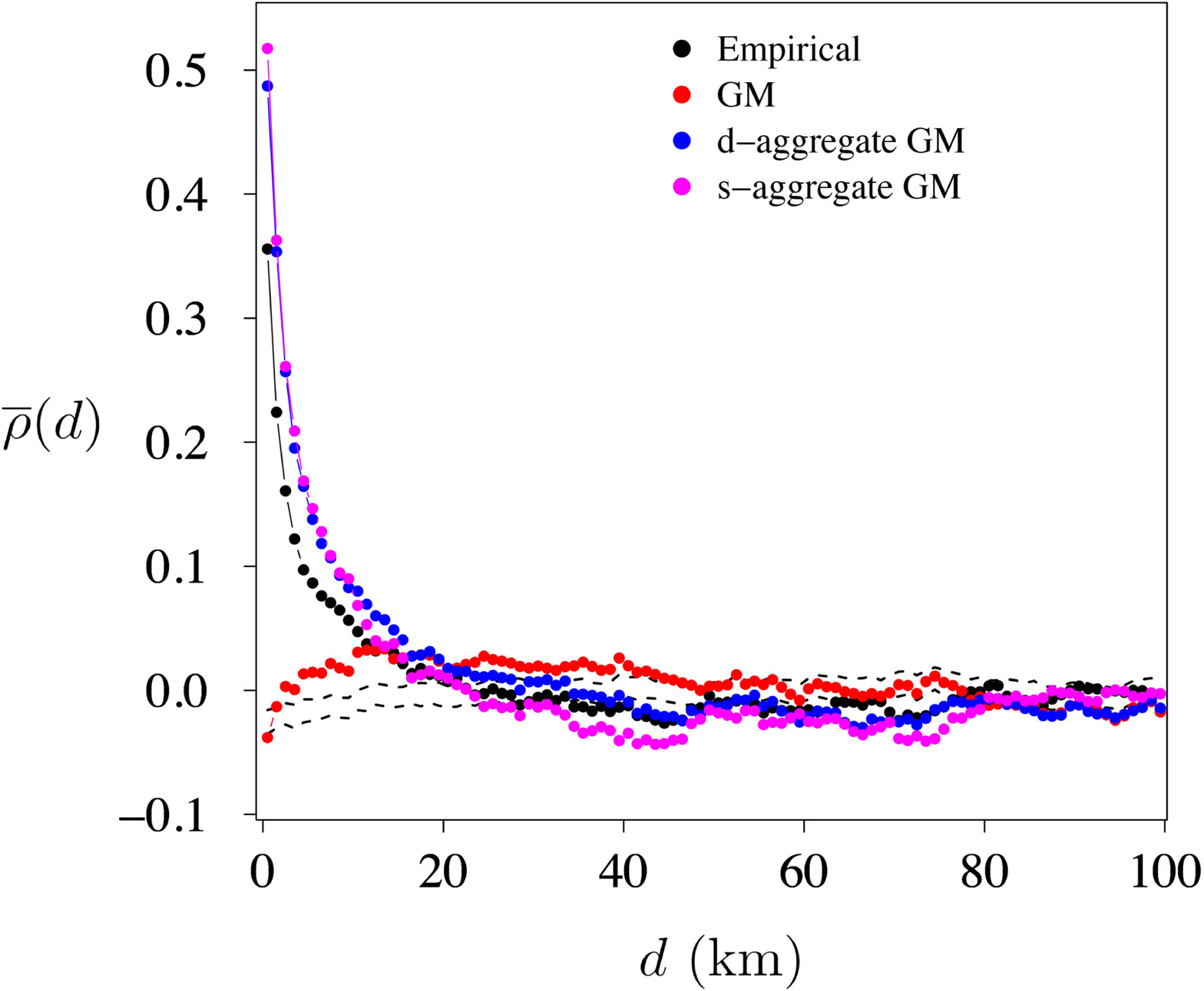}
\end{center}
\caption{Dependence of the population growth rate in a cell on the population density at distance $\textcolor{black}{d}$, $\overline{\rho}(\textcolor{black}{d})$, calculated from the empirical and numerical data between 2000 and 2005. We set $\alpha=0.8$, $\beta=0.4$ \textcolor{black}{and $\gamma=1$ for the GM; $\alpha=0.8$, $\beta=0.4$, $\gamma=1$ and $d_{ag}=0.65$ km for the d-aggregate GM; $\alpha=0.4$, $\beta=1.2$, $\gamma=1$ and $d_{ag}=0.65$ km for the s-aggregate GM.} The ranges indicated by the dashed lines represent 95\% CIs.}
\label{rho12}
\end{figure}

\newpage
\begin{figure}[H]
\begin{center}
\includegraphics[width=160mm]{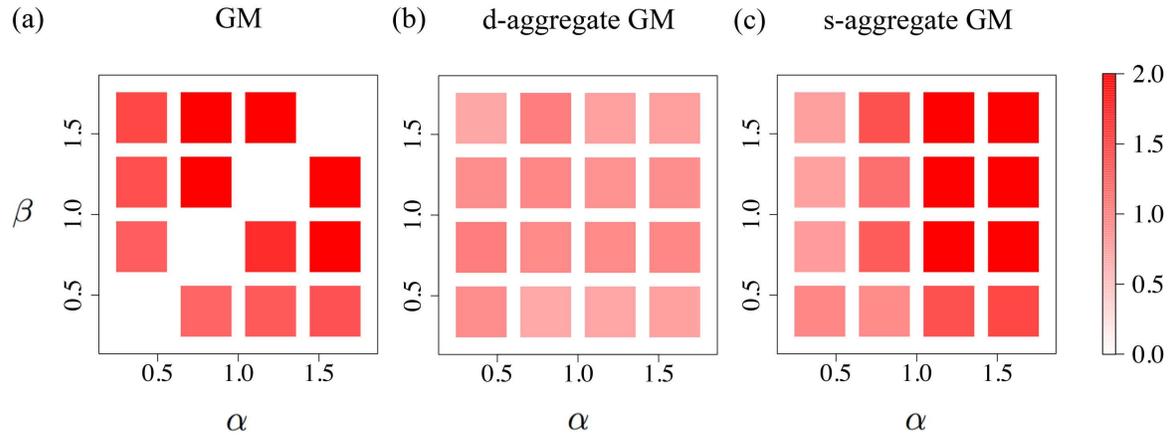}
\end{center}
\caption{The discrepancy of the GM, d-aggregate GM and s-aggregate GM from the empirical data, calculated for the population change between 2000 and 2005. (a) GM. (b) d-aggregate GM. (c) s-aggregate GM. The diagonal in (a) is blank because the in-flow and out-flow are equal when $\alpha = \beta$ in the GM, resulting in no population change. We set $\gamma = 1$ \textcolor{black}{and $\textcolor{black}{d}_{ag}=0.65$ km}.}
\label{accuracy_m12}
\end{figure}

\newpage
\begin{figure}[H]
\begin{center}
\includegraphics[width=160mm]{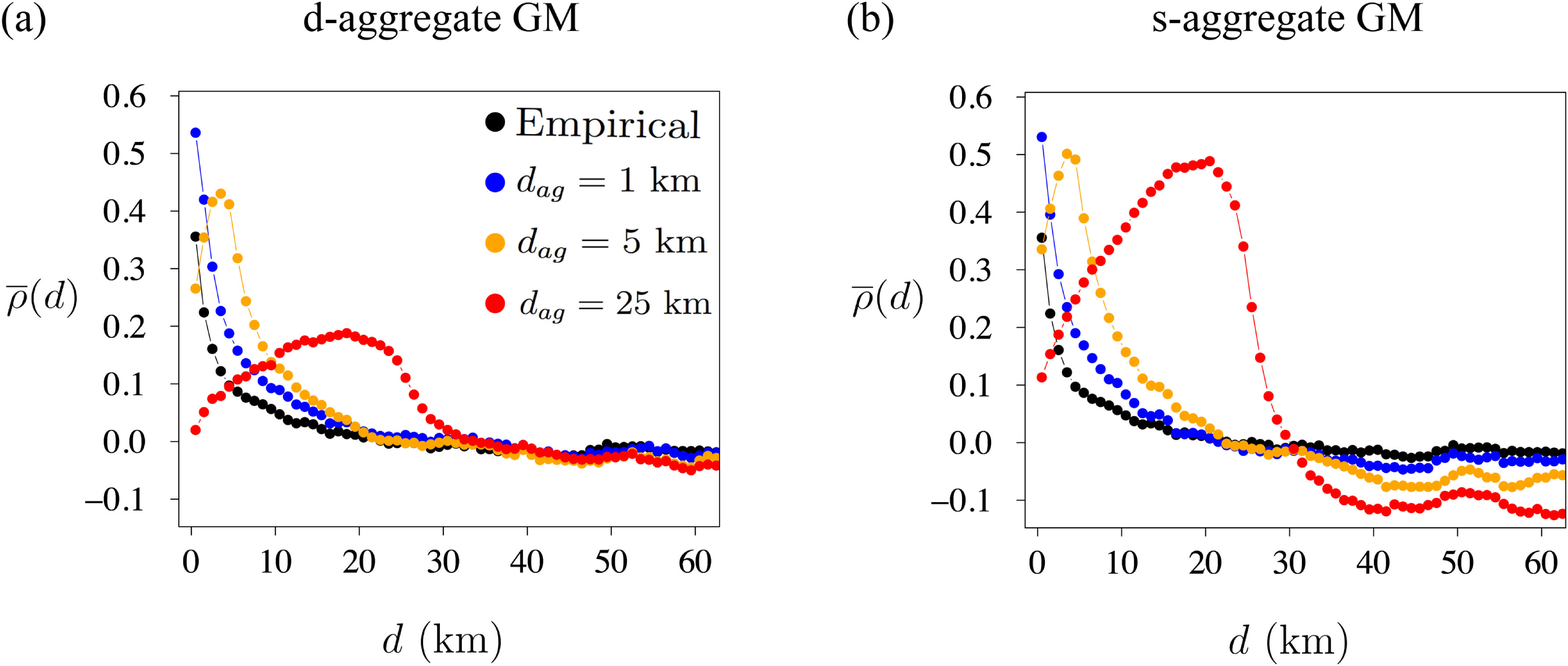}
\end{center}
\caption{Dependence of the population growth rate in a cell on the population density at distance $\textcolor{black}{d}$, $\overline{\rho}(\textcolor{black}{d})$, calculated for the population change between 2000 and 2005. We varied the values of $\textcolor{black}{d}_{\rm ag}$. (a) d-aggregate GM. \textcolor{black}{We set $\alpha=0.8$, $\beta=0.4$ and $\gamma = 1.0$.} (b) s-aggregate GM. \textcolor{black}{We set $\alpha=0.4$, $\beta=1.2$ and $\gamma = 1.0$.}} 
\label{rho_aggregate_range12}
\end{figure}

\newpage
\begin{figure}[H]
\begin{center}
\includegraphics[width=130mm]{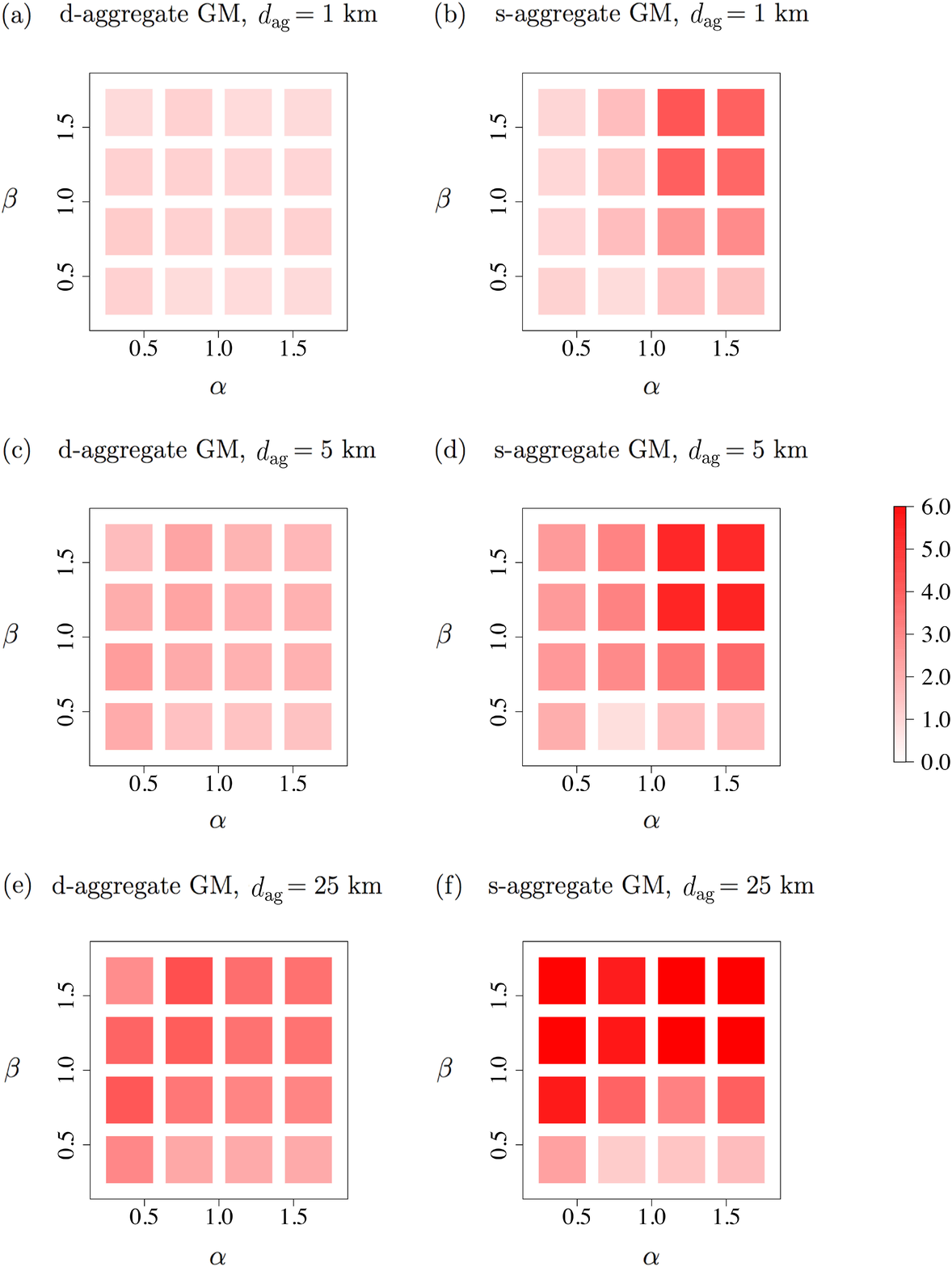}
\end{center}
\caption{The discrepancy of the d-aggregate GM and s-aggregate GM from the empirical data for the population change between 2000 and 2005. (a) d-aggregate GM, $\textcolor{black}{d}_{\rm ag}=1$ km. (b) s-aggregate GM, $\textcolor{black}{d}_{\rm ag}=1$ km. (c) d-aggregate GM, $\textcolor{black}{d}_{\rm ag}=5$ km. (d) s-aggregate GM, $\textcolor{black}{d}_{\rm ag}=5$ km. (e) d-aggregate GM, $\textcolor{black}{d}_{\rm ag}=25$ km. (f) s-aggregate GM, $\textcolor{black}{d}_{\rm ag}=25$ km.}
\label{accuracy_range12}
\end{figure}

\def\theequation{B \arabic{equation}}
\makeatother
\setcounter{equation}{0}

\newpage
\begin{figure}[H]
\begin{center}
\includegraphics[width=160mm]{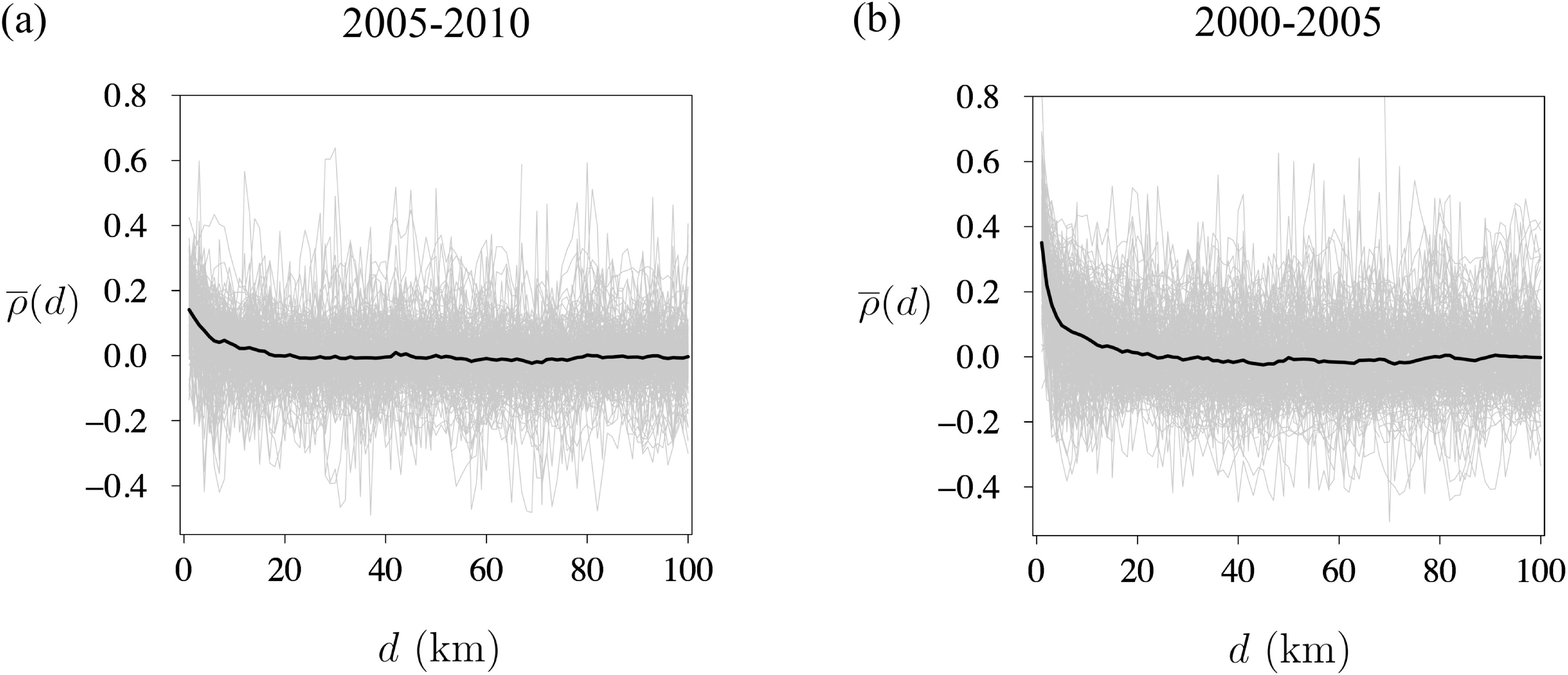}
\end{center}
\caption{Dependence of the \textcolor{black}{population} growth rate in a cell on the population density at distance $\textcolor{black}{d}$, $\rho_{k}(\textcolor{black}{d})$, calculated from the empirical data. A thin line represents $\rho_{k}(\textcolor{black}{d})$ for a region \textcolor{black}{of size 50 km $\times$ 50 km. The results for the different regions are superposed on top of each other}. \textcolor{black}{The thick lines represent} $\overline{\rho}(\textcolor{black}{d})$, which is the average of $\rho_{k}(\textcolor{black}{d})$ \textcolor{black}{over all the regions. The thick lines in (a) and (b) are the same as the lines with the black circles shown in Figs.~4 and~10, respectively.} (a) \textcolor{black}{($t_1$, $t_2$) = (2005, 2010)}. (b) \textcolor{black}{($t_1$, $t_2$) = (2000, 2005)}. }
\label{rho_subregions}
\end{figure}

\newpage
\begin{figure}[H]
\begin{center}
\includegraphics[width=160mm]{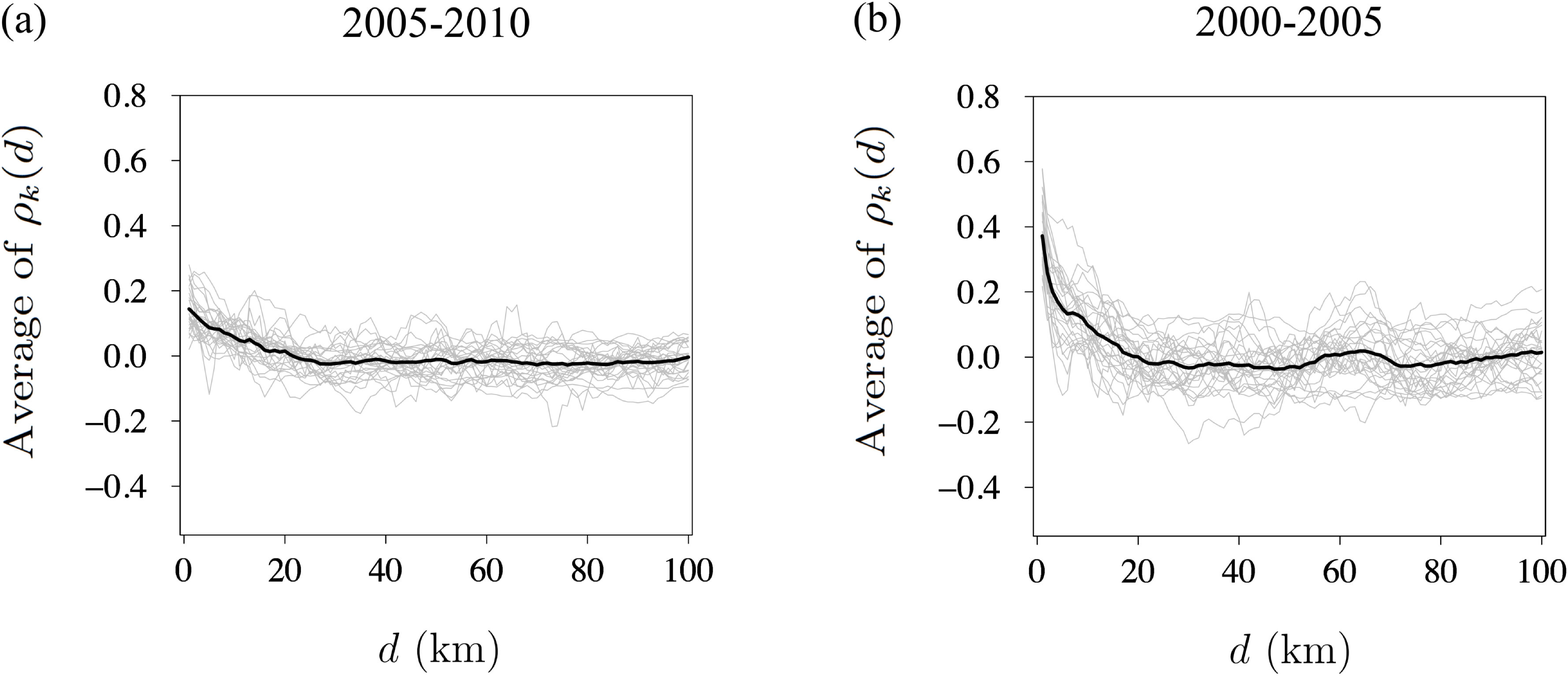}
\end{center}
\caption{\textcolor{black}{Dependence of the population growth rate in a cell on the population density at distance $d$, $\rho_{k}(d)$, calculated from the empirical data. We calculated $\rho_{k}(d)$ for regions $k$ such that all cells within 30 km from any cell in region $k$ do not contain sea. A thin line represents $\rho_{k}(d)$ for such a 50 km $\times$ 50 km region. The results for the different regions are superposed on top of each other. The thick lines represent the average of $\rho_{k}(d)$ over all the regions satisfying the aforementioned criterion.  (a) ($t_1$, $t_2$) = (2005, 2010). (b) ($t_1$, $t_2$) = (2000, 2005)}. }
\label{rho_30km}
\end{figure}

\def\theequation{C \arabic{equation}}
\makeatother
\setcounter{equation}{0}

\newpage
\begin{figure}[H]
\begin{center}
\includegraphics[width=160mm]{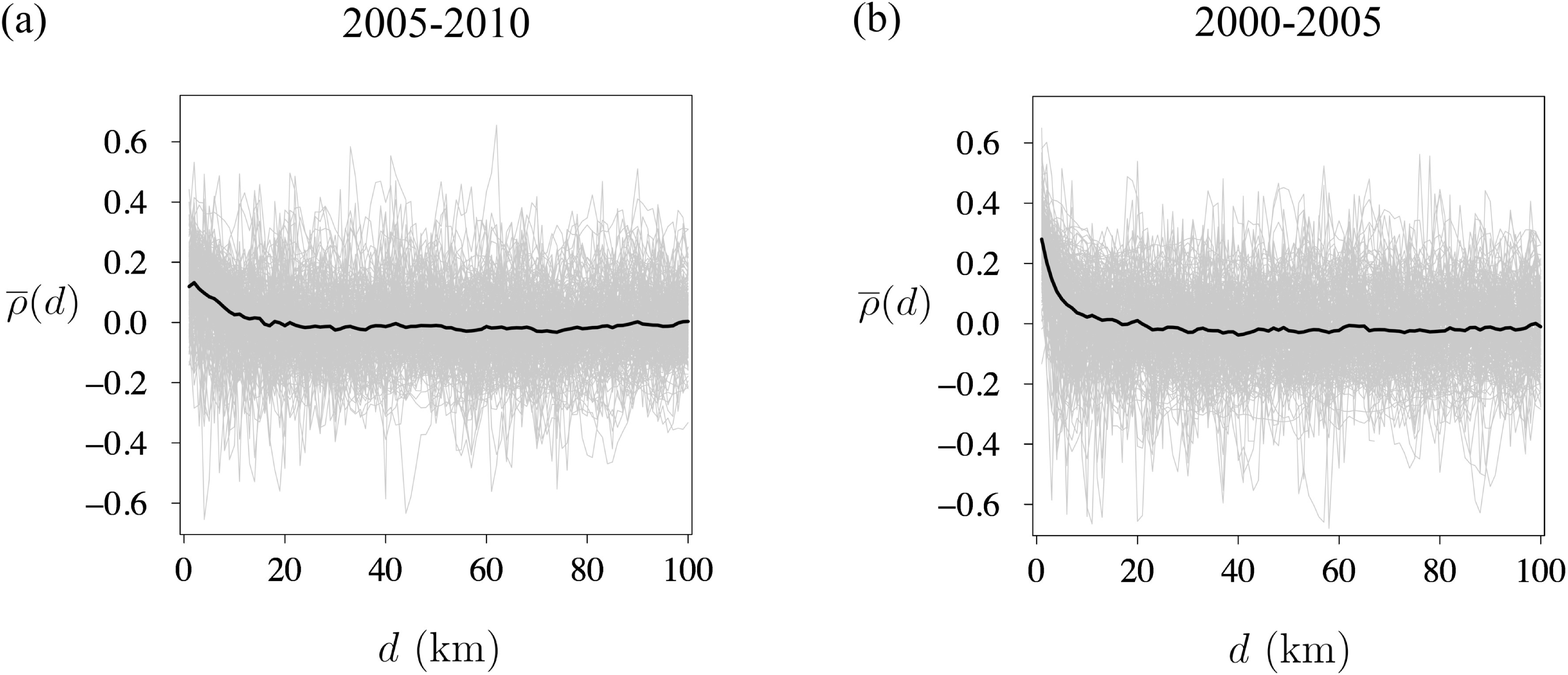}
\end{center}
\caption{Dependence of the growth rate in a cell on the population density at distance $\textcolor{black}{d}$, $\rho_{k}(\textcolor{black}{d})$, when \textcolor{black}{only the} cells whose \textcolor{black}{number of inhabitants is} greater than 100 \textcolor{black}{are considered}. A thin line \textcolor{black}{represents} $\rho_{k}(\textcolor{black}{d})$ for a region of size 50 km $\times$ 50 km. The thick lines represent $\overline{\rho}(\textcolor{black}{d})$, which is the average of $\rho_{k}(\textcolor{black}{d})$ \textcolor{black}{over all the regions}. (a) \textcolor{black}{($t_1$, $t_2$) = (2005, 2010)}. (b) \textcolor{black}{($t_1$, $t_2$) = (2000, 2005)}. }
\label{rho_over100}
\end{figure}

\def\theequation{D \arabic{equation}}
\makeatletter
\makeatother
\setcounter{equation}{0}

\newpage
\begin{figure}[H]
\begin{center}
\includegraphics[width=160mm]{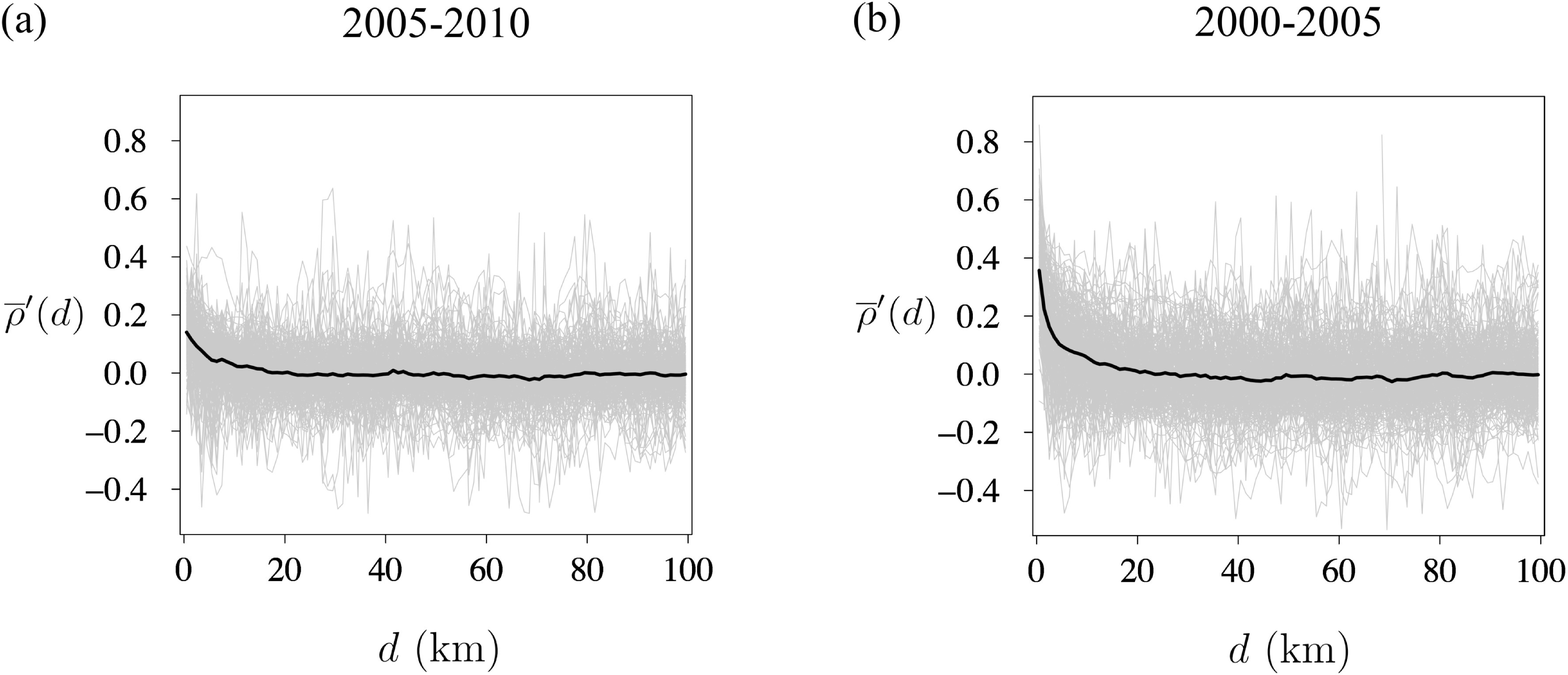}
\end{center}
\caption{Dependence of the growth rate in a cell on the population density at distance $\textcolor{black}{d}$, $\rho^{\prime}_{k}(\textcolor{black}{d})$, controlling for the population size of a focal cell. A thin line \textcolor{black}{represents} $\rho^{\prime}_{k}(\textcolor{black}{d})$ for a region. The thick lines \textcolor{black}{represent} $\overline{\rho}^{\prime}(\textcolor{black}{d})$, which is the average of $\rho^{\prime}_{k}(\textcolor{black}{d})$ \textcolor{black}{over all the regions}. (a) \textcolor{black}{($t_1$, $t_2$) = (2005, 2010)}. (b) \textcolor{black}{($t_1$, $t_2$) = (2000, 2005)}.}
\label{partial_cor}
\end{figure}

\def\theequation{E \arabic{equation}}
\renewcommand{\figurename}{Figure}
\makeatletter
\makeatother
\setcounter{equation}{0}

\newpage
\begin{figure}[H]
\begin{center}
\includegraphics[width=160mm]{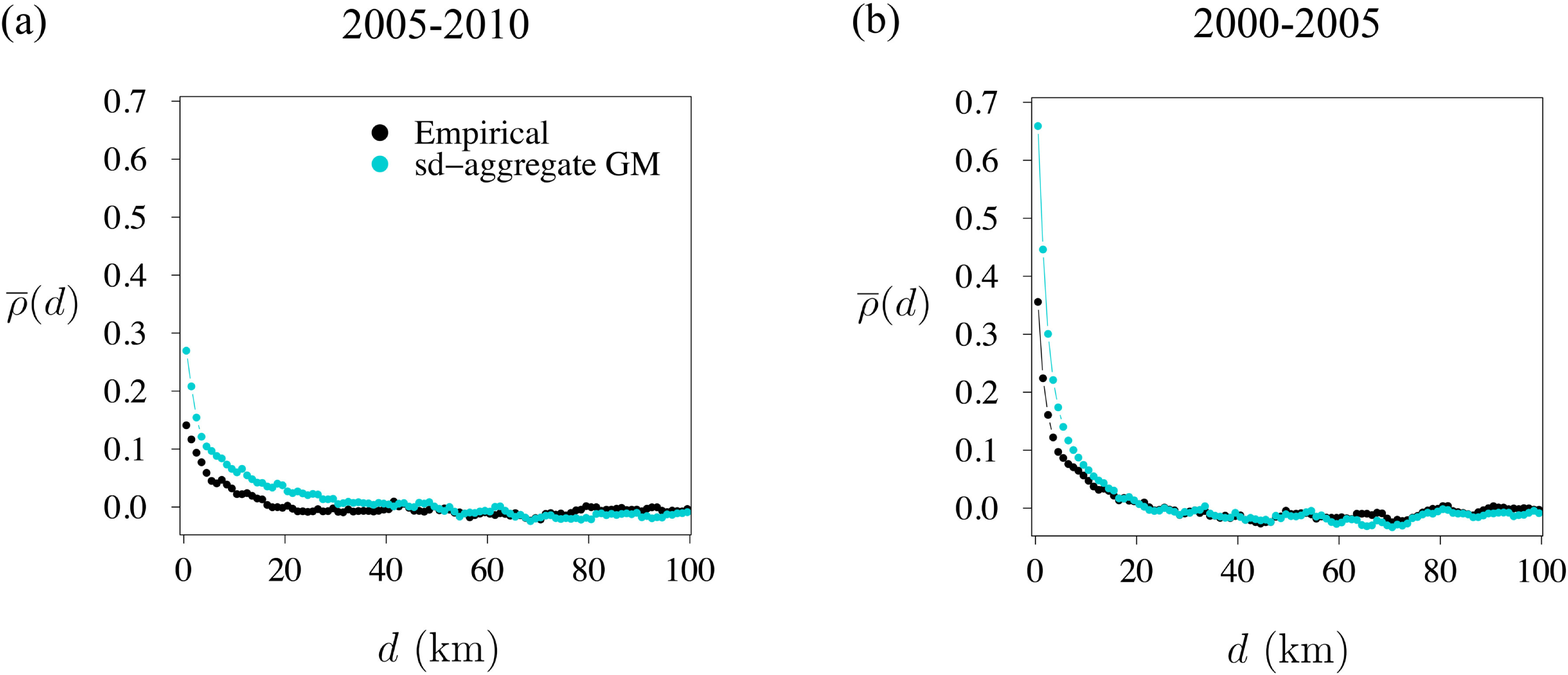}
\end{center}
\caption{Dependence of the population growth rate in a cell on the population density at distance $\textcolor{black}{d}$, $\overline{\rho}(\textcolor{black}{d})$, calculated from the empirical data and the numerical data generated from the sd-aggregate GM. (a) \textcolor{black}{($t_1$, $t_2$) = (2005, 2010)}. \textcolor{black}{We set $\alpha=0.4$, $\beta=1.6$, $\gamma=1$ and $d_{ag}=0.65$ km.} (b) \textcolor{black}{($t_1$, $t_2$) = (2000, 2005)}. \textcolor{black}{We set $\alpha=0.4$, $\beta=0.4$, $\gamma=1$ and $d_{ag}=0.65$ km.}}
\label{rho_sd-aggregate}
\end{figure}

\newpage
\begin{figure}[H]
\begin{center}
\includegraphics[width=160mm]{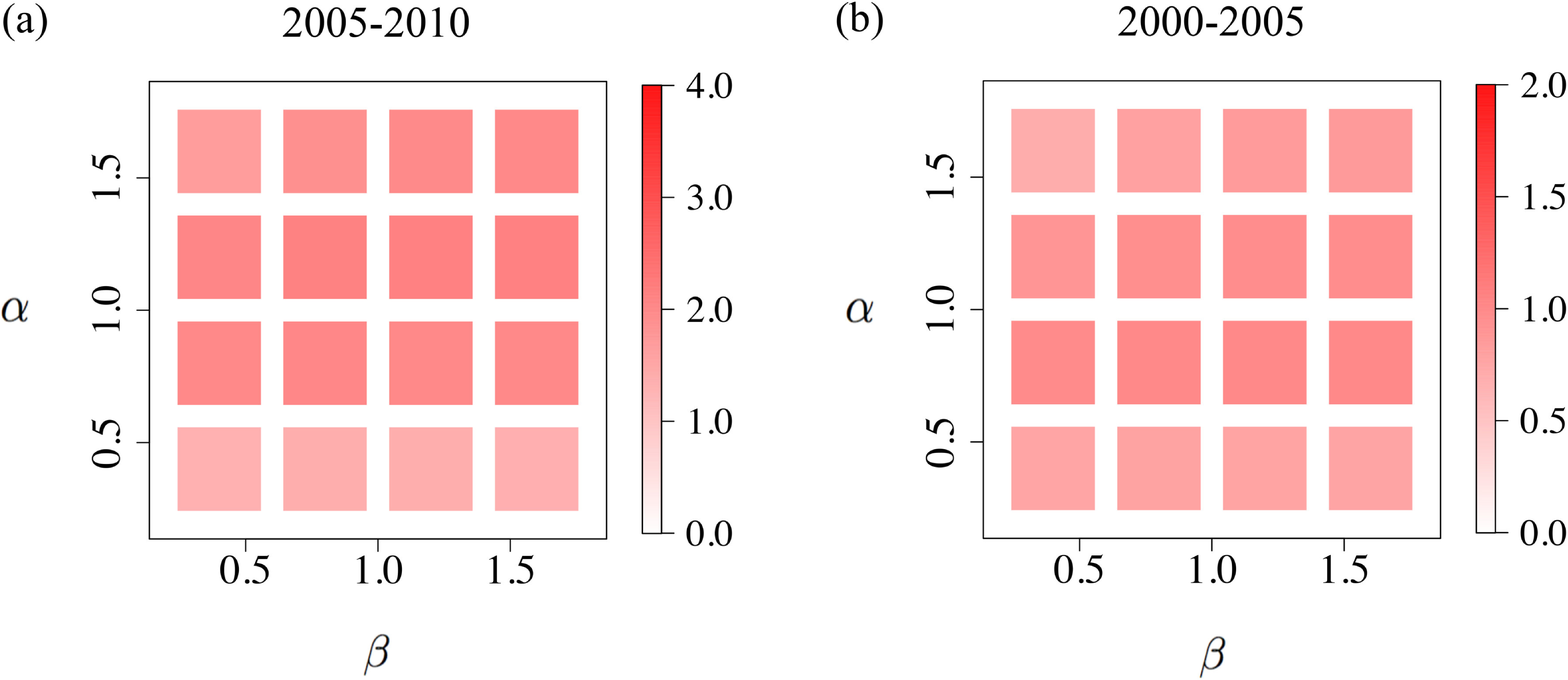}
\end{center}
\caption{The discrepancy of the sd-aggregate GM from the empirical data. (a) \textcolor{black}{($t_1$, $t_2$) = (2005, 2010)}. (b) \textcolor{black}{($t_1$, $t_2$) = (2000, 2005)}. We set $\gamma = 1$ \textcolor{black}{and $\textcolor{black}{d}_{ag}=0.65$ km}.}
\label{accuracy_sd-aggregate}
\end{figure}

\newpage
\begin{figure}[H]
\begin{center}
\includegraphics[width=160mm]{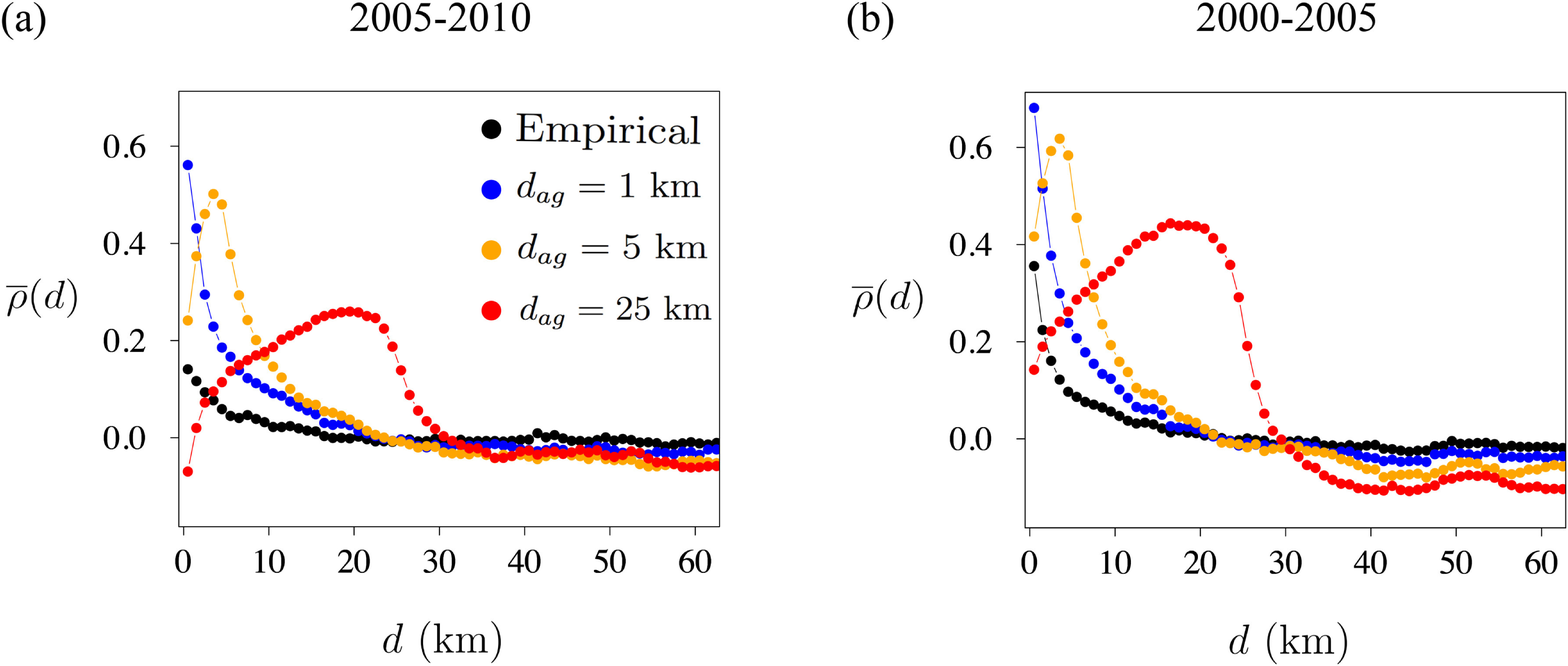}
\end{center}
\caption{Dependence of the population growth rate in a cell on the population density at distance $\textcolor{black}{d}$, $\overline{\rho}(\textcolor{black}{d})$, calculated for the sd-aggregate GM for different values of $\textcolor{black}{d}_{\rm ag}$. (a) \textcolor{black}{($t_1$, $t_2$) = (2005, 2010)}. \textcolor{black}{We set $\alpha=0.4$, $\beta=1.6$ and $\gamma = 1.0$.} (b) \textcolor{black}{($t_1$, $t_2$) = (2000, 2005)}. \textcolor{black}{We set $\alpha=0.4$, $\beta=0.4$ and $\gamma = 1.0$.}}
\label{aggregate_range_sd-aggregate}
\end{figure}

\newpage
\begin{figure}[H]
\begin{center}
\includegraphics[width=130mm]{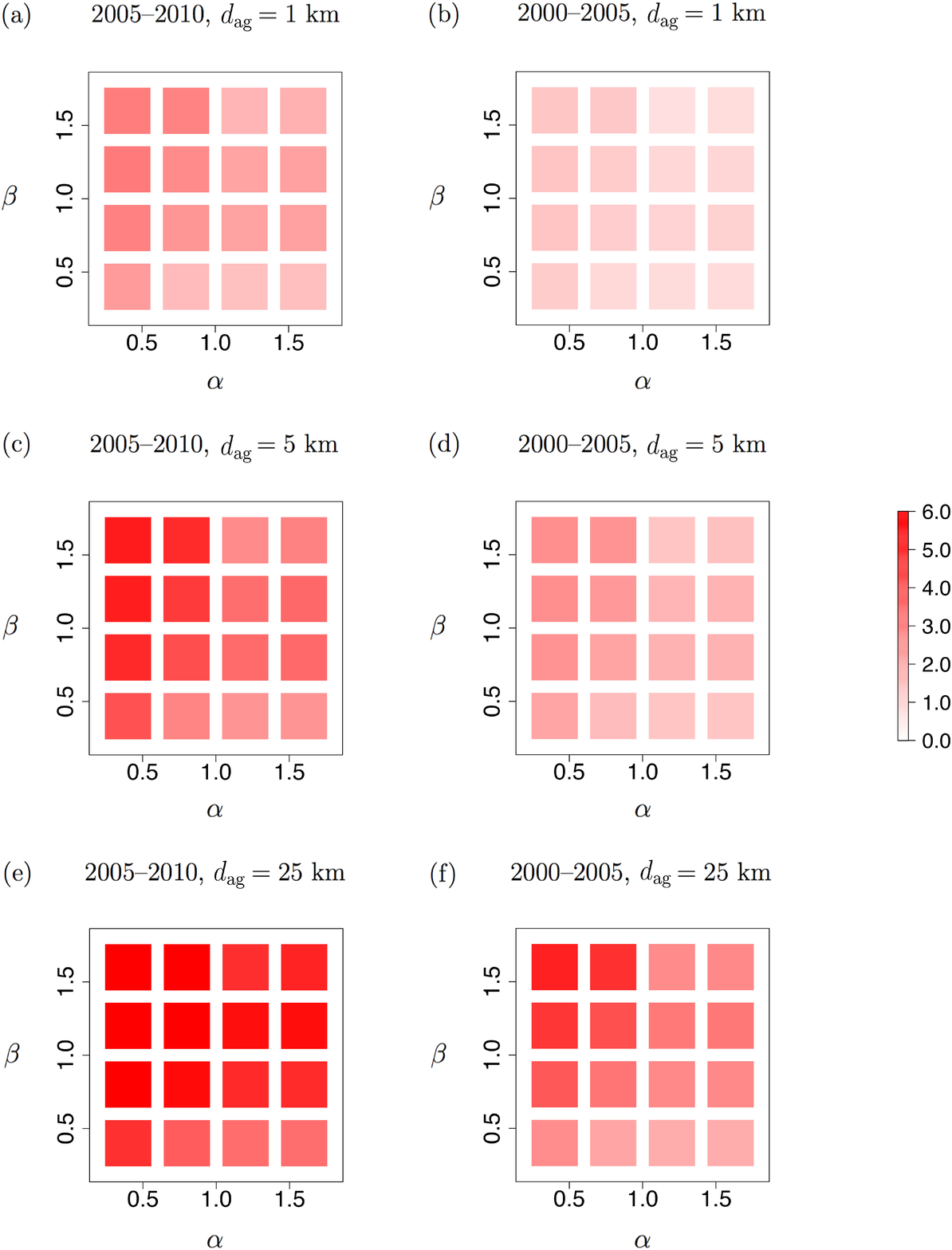}
\end{center}
\caption{The discrepancy of the sd-aggregate GM from the empirical data. (a) \textcolor{black}{($t_1$, $t_2$) = (2005, 2010)}, $\textcolor{black}{d}_{\rm ag}=1$ km. (b) \textcolor{black}{($t_1$, $t_2$) = (2000, 2005)}, $\textcolor{black}{d}_{\rm ag}=1$ km. (c) \textcolor{black}{($t_1$, $t_2$) = (2005, 2010)}, $\textcolor{black}{d}_{\rm ag}=5$ km. (d) \textcolor{black}{($t_1$, $t_2$) = (2000, 2005)}, $\textcolor{black}{d}_{\rm ag} = 5$ km. (e) \textcolor{black}{($t_1$, $t_2$) = (2005, 2010)}, $\textcolor{black}{d}_{\rm ag} = 25$ km. (f) \textcolor{black}{($t_1$, $t_2$) = (2000, 2005)}, $\textcolor{black}{d}_{\rm ag} = 25$ km.}
\label{accuracy_range_sd-aggregate}
\end{figure}

\newpage
\begin{figure}[H]
\begin{center}
\includegraphics[width=150mm]{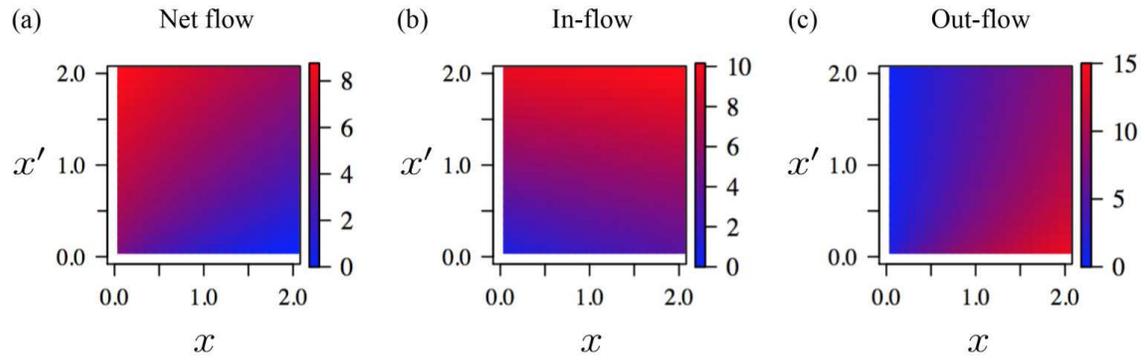}
\end{center}
\caption{The net flow, in-flow, and out-flow for the sd-aggregate GM in \textcolor{black}{the} one-dimensional model. \textcolor{black}{We set} $G = (1/3)^{\alpha + \beta -1}$\textcolor{black}{,} $\alpha = 0.4$, $\beta = 0.6$ and $\gamma = 1.0$. (a) Net flow. (b) In-flow. (c) Out-flow.}
\label{flow_sd-aggregate}
\end{figure}

\newpage
\section*{Tables}
\def\theequation{}
\renewcommand{\figurename}{Figure}
\makeatletter
\renewcommand{\thefigure}{\arabic{figure}}
\renewcommand{\thetable}{\arabic{table}}
\makeatother
\setcounter{table}{0}
\setcounter{figure}{0}
\setcounter{equation}{0}

\begin{table}[htb]
\begin{center}
    \caption{Statistics of the data set}
  \begin{tabular}{cccc}
    \hline
    Year & 2000 & 2005 & 2010 \\ \hline
    Total population & 126,925,843 & 127,767,994 & 128,057,352 \\
    Average number of inhabitants in a cell & \textcolor{black}{249.48} & \textcolor{black}{251.13} & \textcolor{black}{251.70} \\
    Median number of inhabitants in a cell & 33 & 41 & 38 \\
    Number of populated cells & 308,418 & 482,181 & 477,172 \\ \hline
  \end{tabular}
  \label{tab:summary}
  \end{center}
\end{table}

\begin{table}[H]
\begin{center}
\caption{}
\begin{tabular}{|c|ccccc|}
\hline
Prefecture & Births & Deaths & In-flow & Out-flow & RC \\ \hline
Hokkaido & 206,018 & 258,620 & 1,400,785 & 1,486,704 & 0.861 \\
Aomori & 50,846 & 75,402 & 210,797 & 253,252 & 0.786 \\
Iwate & 51,415 & 74,612 & 208,183 & 237,908 & 0.780 \\
Miyagi & 98,143 & 101,836 & 569,239 & 590,600 & 0.853 \\
Akita & 37,200 & 68,304 & 132,922 & 161,071 & 0.736 \\
Yamagata & 45,724 & 67,185 & 162,073 & 182,237 & 0.753 \\
Fukushima & 84,844 & 106,394 & 299,166 & 337,621 & 0.769 \\
Ibaraki & 123,337 & 134,098 & 539,995 & 543,333 & 0.808 \\
Tochigi & 85,950 & 91,337 & 338,813 & 343,879 & 0.794 \\
Gunma & 84,469 & 93,531 & 315,194 & 323,801 & 0.782 \\
Saitama & 302,796 & 251,786 & 1,673,676 & 1,612,124 & 0.856 \\
Chiba & 259,317 & 230,495 & 1,573,082 & 1,479,729 & 0.862 \\
Tokyo & 518,801 & 481,388 & 4,228,697 & 3,855,587 & 0.890 \\
Kanagawa & 393,305 & 307,305 & 2,434,444 & 2,297,378 & 0.871 \\
Nigata & 92,614 & 123,745 & 274,628 & 301,798 & 0.727\\
Toyama & 43,760 & 56,352 & 134,116 & 142,212 & 0.734 \\
Ishikawa & 50,547 & 52,753 & 181,465 & 190,238 & 0.783 \\
Fukui & 35,888 & 39,657 & 100,717 & 111,694 & 0.738 \\
Yamanashi & 34,806 & 42,633 & 155,103 & 166,314 & 0.806 \\
Nagano & 91,097 & 109,115 & 369,322 & 389,224 & 0.791 \\
Gifu & 88,156 & 95,085 & 323,422 & 344,227 & 0.785 \\
Shizuoka & 163,151 & 165,452 & 702,346 & 710,330 & 0.811 \\
Aichi & 347,947 & 269,444 & 1,679,203 & 1,602,590 & 0.842 \\
Mie & 78,283 & 87,013 & 305,184 & 310,744 & 0.788 \\ \hline
\end{tabular}
\label{tab:migration}
\end{center}
\end{table}

\def\theequation{}
\renewcommand{\figurename}{Figure}
\makeatletter
\renewcommand{\thefigure}{\arabic{figure}}
\renewcommand{\thetable}{\arabic{table}}
\makeatother
\setcounter{table}{1}
\setcounter{figure}{0}
\setcounter{equation}{0}

\begin{table}[H]
\begin{center}
 \begin{tabular}{|c|ccccc|}
\hline 
Prefecture & Births & Deaths & In-flow & Out-flow & RC \\ \hline
Shiga & 66,776 & 54,067 & 271,852 & 260,071 & 0.815 \\
Kyoto & 108,329 & 113,508 & 605,810 & 623,003 & 0.847 \\
Osaka & 383,123 & 354,039 & 2,038,886 & 2,050,292 & 0.847 \\
Hyogo & 241,991 & 239,518 & 1,092,350 & 1,099,138 & 0.820 \\
Nara & 55,617 & 59,986 & 248,748 & 271,018 & 0.818 \\
Wakayama & 38,888 & 57,061 & 134,100 & 153,523 & 0.750 \\
Tottori & 24,973 & 32,708 & 89,781 & 101,313 & 0.768 \\
Shimane & 28,992 & 43,489 & 110,090 & 122,946 & 0.763 \\
Okayama & 84,959 & 93,940 & 306,532 & 316,719 & 0.777 \\
Hiroshima & 127,862 & 131,562 & 601,952 & 613,915 & 0.824 \\
Yamaguchi & 57,910 & 83,418 & 248,565 & 266,357 & 0.785 \\
Tokushima & 30,023 & 43,472 & 122,127 & 133,197 & 0.776 \\
Kagawa & 42,961 & 52,403 & 173,267 & 182,174 & 0.788 \\
Ehime & 57,940 & 76,292 & 213,144 & 231,838 & 0.768 \\
Kochi & 28,744 & 45,893 & 122,876 & 139,076 & 0.778 \\
Fukuoka & 228,884 & 220,437 & 1,308,177 & 1,312,289 & 0.854 \\
Saga & 38,216 & 43,612 & 151,729 & 163,207 & 0.794 \\
Nagasaki & 60,771 & 76,205 & 264,659 & 306,307 & 0.807 \\
Kumamoto & 80,457 & 91,365 & 333,680 & 351,123 & 0.799 \\
Oita & 50,366 & 61,697 & 208,513 & 216,427 & 0.791 \\
Miyazaki & 50,795 & 57,628 & 226,740 & 244,558 & 0.813 \\
Kagoshima & 75,607 & 96,695 & 369,723 & 399,374 & 0.817 \\
Okinawa & 82,886 & 47,009 & 368,805 & 373,402 & 0.851 \\ \hline
\end{tabular}
\label{tab:migration2}
\caption{The number of births, deaths, in-coming inhabitants and out-going inhabitants in the 47 prefectures in Japan between 2005 and 2009. The relative contribution of migration to demographic dynamics, denoted by RC in the table, is defined by ( in-flow $+$ out-flow )/( the number of births $+$  the number of deaths $+$ in-flow $+$ out-flow). The average of RC over the 47 prefectures is 0.801. Data were obtained from Refs.~\cite{JGDC2006, JGDC2007, JGDC2008, JGDC2009, JGDC2010}.}
\end{center}
\end{table}


\newpage
\begin{thebibliography}{10}
\expandafter\ifx\csname urlstyle\endcsname\relax
  \providecommand{\doi}[1]{doi:\discretionary{}{}{}#1}\else
  \providecommand{\doi}{doi:\discretionary{}{}{}\begingroup
  \urlstyle{rm}\Url}\fi

\bibitem{Pallagst2007}
Pallagst K \emph{et~al.} 2009 Shrinking cities in the United States of
  America: three cases, three planning stories.
\newblock In \emph{The future of shrinking cities: Problems, patterns and
  strategies of urban transformation in a global context} (eds K Pallagst, et al.), pp. 81--88. 
  Berkeley, CA: Institute of Urban and Regional Development, University of
California.

\bibitem{Pallagst2013}
Pallagst K, Wiechmann T \& Martinez-Fernandez C. 2013 \emph{Shrinking
  cities: international perspectives and policy implications}.
\newblock Abingdon, UK: Routledge.

\bibitem{Hara2014}
Hara T. 2014 \emph{A shrinking society: post-demographic transition in
  Japan}.
\newblock Tokyo, Japan: Springer.

\bibitem{Lewis1982}
Lewis GJ. 1982 \emph{Human migration: a geographical perspective}.
\newblock Kent, UK: Croom Helm.

\bibitem{Kahley1991}
Kahley WJ. 1991 Population migration in the United States: a survey of
  research.
\newblock \emph{Econ. Rev.} \textbf{76}, 12--21.

\bibitem{Beine2014}
Beine M, Bertoli S \& Moraga JF-H. 2014 A practitioners' guide to
  gravity models of international migration.
\newblock \emph{FEDEA Working Paper 2014-03} (\doi{10.1111/twec.12265}).

\bibitem{Stouffer1940}
Stouffer SA. 1940 Intervening opportunities: a theory relating mobility and
  distance.
\newblock \emph{Am. Sociol. Rev.} \textbf{5}, 845--867.

\bibitem{Zipf1946}
Zipf GK. 1946 {The $P_1$$P_2$/D hypothesis: on the intercity movement of
  persons}.
\newblock \emph{Am. Sociol. Rev.} \textbf{11}, 677--686.

\bibitem{Cohen2008}
Cohen JE, Roig M, Reuman DC \& GoGwilt C. 2008 International
  migration beyond gravity: a statistical model for use in population
  projections.
\newblock \emph{Proc. Natl. Acad. Sci. USA} \textbf{105}, 15269--15274.
\newblock (\doi{10.1073/pnas.0808185105}).

\bibitem{Batty2013}
Batty M. 2013 \emph{The new science of cities}.
\newblock Cambridge, MA: MIT Press.

\bibitem{Simini2012}
Simini F, Gonz{\'a}lez MC, Maritan A \& Barab{\'a}si A-L. 2012 A
  universal model for mobility and migration patterns.
\newblock \emph{Nature} \textbf{484}, 96--100.
\newblock (\doi{10.1038/nature10856}).

\bibitem{Simini2013}
Simini F, Maritan A \& N{\'e}da Z. 2013 Human mobility in a continuum
  approach.
\newblock \emph{PLOS ONE} \textbf{8}, e60069.
\newblock (\doi{10.1371/journal.pone.0060069}).

\bibitem{Barthelemy2016}
Barthelemy M. 2016 \emph{The structure and dynamics of cities: urban data
  analysis and theoretical modeling}.
\newblock Cambridge, UK: Cambridge University Press.

\bibitem{Anderson2011}
Anderson JE. 2011 The gravity model.
\newblock \emph{Annu. Rev. Econ.} \textbf{3}, 133--160.
\newblock (\doi{10.1146/annurev-economics-111809-125114}).

\bibitem{Rodrigue2013}
Rodrigue JP, Comtois C \& Slack B. 2013 \emph{The geography of
  transport systems}.
\newblock Abingdon, UK: Routledge.

\bibitem{Karemera2000}
Karemera D, Oguledo VI \& Davis B. 2000 A gravity model analysis of
  international migration to North America.
\newblock \emph{Appl. Econ.} \textbf{32}, 1745--1755.
\newblock (\doi{10.1080/000368400421093}).

\bibitem{Fagiolo2013}
Fagiolo G \& Mastrorillo M. 2013 International migration network: topology
  and modeling.
\newblock \emph{Phys. Rev. E} \textbf{88}, 012812.
\newblock (\doi{10.1103/PhysRevE.88.012812}).

\bibitem{Bhattacharya2008}
Bhattacharya K, Mukherjee G, Saram{\"a}ki J, Kaski K \& Manna SS.
 2008 The international trade network: weighted network analysis and
  modelling.
\newblock \emph{J. Stat. Mech.} \textbf{2008}, P02002.
\newblock (\doi{10.1088/1742-5468/2008/02/P02002}).

\bibitem{Kepaptsoglou2010}
Kepaptsoglou K, Karlaftis MG \& Tsamboulas D. 2010 The gravity model
  specification for modeling international trade flows and free trade agreement
  effects: a 10-year review of empirical studies.
\newblock \emph{Open Econ. J.} \textbf{3}, 1--13.
\newblock (\doi{10.2174/1874919401003010001}).

\bibitem{Lambiotte2008}
Lambiotte R, Blondel VD, De~Kerchove C, Huens E, Prieur C, Smoreda
 Z \& Van~Dooren P. 2008 Geographical dispersal of mobile communication
  networks.
\newblock \emph{Physica A} \textbf{387}, 5317--5325.
\newblock (\doi{10.1016/j.physa.2008.05.014}).

\bibitem{Krings2009}
Krings G, Calabrese F, Ratti C \& Blondel VD. 2009 Urban gravity: a
  model for inter-city telecommunication flows.
\newblock \emph{J. Stat. Mech.} \textbf{2009}, L07003.
\newblock (\doi{10.1088/1742-5468/2009/07/L07003}).

\bibitem{Akwawua2016}
Akwawua S \& Pooler JA. 2000 An intervening opportunities model of US
  interstate migration flows.
\emph{Geo. Res. Forum}, \textbf{20}, 33--51.

\bibitem{Openshaw1984}
Openshaw S. 1984 Ecological fallacies and the analysis of areal census data.
\newblock \emph{Environ. Plan. A} \textbf{16}, 17--31.
\newblock (\doi{10.1068/a160017}).

\bibitem{Openshaw1977}
Openshaw S. 1977 Optimal zoning systems for spatial interaction models.
\newblock \emph{Environ. Plan. A} \textbf{9}, 169--184.
\newblock (\doi{10.1068/a090169s}).

\bibitem{Broadbent1970}
Broadbent TA. 1970 Notes on the design of operational models.
\newblock \emph{Environ. Plan. A} \textbf{2}, 469--476.

\bibitem{Batty1974}
Batty M. 1974 Spatial entropy.
\newblock \emph{Geogr. Anal.} \textbf{6}, 1--31.
\newblock (\doi{10.1111/j.1538-4632.1974.tb01014.x}).

\bibitem{Batty1976}
Batty M. 1976 Entropy in spatial aggregation.
\newblock \emph{Geogr. Anal.} \textbf{8}, 1--21.
\newblock (\doi{10.1111/j.1538-4632.1976.tb00525.x}).

\bibitem{Masser1977}
Masser I \& Brown PJ. 1977 Spatial representation and spatial
  interaction.
\newblock \emph{Pap. Reg. Sci.}, 38, 71--92.
\newblock (\doi{10.1111/j.1435-5597.1977.tb00992.x}).

\bibitem{estatjp}
{Portal Site of Official Statistics of Japan}.
\newblock http://e-stat.go.jp/SG2/eStatGIS/page/download.html.
\newblock Accessed 2015 March 17.
\bibitem{gaiyo}
{Somu-sho Tokei-kyoku ni okeru Chiiki Mesh Tokei no sakusei}.
\newblock [in Japanese]
\newblock http://www.stat.go.jp/data/mesh/pdf/gaiyo2.pdf.
\newblock Accessed 2016 December 26.

\bibitem{Gallos2012}
Gallos LK, Barttfeld P, Havlin S, Sigman M \& Makse HA. 2012
  Collective behavior in the spatial spreading of obesity.
\newblock \emph{Sci. Rep.} \textbf{2}, 454.
\newblock (\doi{10.1038/srep00454}).

\bibitem{Rybski2013}
Rybski, D., Ros, A. G.~C. \& Kropp, J.~P., 2013 Distance-weighted city growth.
\newblock \emph{Phys. Rev. E} \textbf{87}, 042114.

\bibitem{Hernando2014}
Hernando A, Hernando R \& Plastino A. 2014 Space--time correlations in
  urban sprawl.
\newblock \emph{J. R. Soc. Interface} \textbf{11}, 20130930.
\newblock (\doi{10.1098/rsif.2013.0930}).

\bibitem{Hernando2015}
Hernando A, Hernando R, Plastino A \& Zambrano E. 2015 Memory-endowed
  US cities and their demographic interactions.
\newblock \emph{J. R. Soc. Interface} \textbf{12}, 20141185.

\bibitem{Michaels2012}
Michaels G, Rauch F \& Redding SJ. 2012 Urbanization and structural
  transformation.
\newblock \emph{Q. J. Econ.} \textbf{127}, 535--586.

\bibitem{Birchenall2016}
Birchenall JA. 2016 Population and development redux.
\newblock \emph{J. Popul. Econ.} \textbf{29}, 627--656.

\bibitem{Desmet2015}
Desmet K \& Rappaport J. 2015 The settlement of the United States,
  1800--2000: the long transition towards Gibrat's law.
\newblock \emph{J. Urban. Econ.} \textbf{98}, 50--68. (\doi{10.1016/j.jue.2015.03.004}).

\bibitem{Rozenfeld2008}
Rozenfeld HD, Rybski D, Andrade JS, Batty M, Stanley HE \&
  Makse HA. 2008 Laws of population growth.
\newblock \emph{Proc. Natl. Acad. Sci. USA} \textbf{105}, 18702--18707.
\newblock (\doi{10.1073/pnas.0807435105}).

\bibitem{Barthelemy2011}
Barth{\'e}lemy M. 2011 Spatial networks.
\newblock \emph{Phys. Rep.} \textbf{499}, 1--101.
\newblock (\doi{10.1016/j.physrep.2010.11.002}).

\bibitem{Rozenfeld2009}
Rozenfeld HD, Rybski D, Gabaix X \& Makse HA. 2011 The area and
  population of cities: new insights from a different perspective on cities.
\newblock \emph{Am. Econ. Rev.} \textbf{101}, 2205--2225.
\newblock (\doi{10.3386/w15409}).

\bibitem{JGDC2006}
{Japan Geographic Data Center}. 2006
 \emph{{Jyumin Kihon Dai-cho jinko yoran Heisei 18 nen ban}}.
\newblock [in Japanese]
\newblock Tokyo, Japan: Japan Statistical Association.

\bibitem{JGDC2007}
{Japan Geographic Data Center}. 2007 
 \emph{{Jyumin Kihon Dai-cho jinko yoran Heisei 19 nen ban}}.
\newblock [in Japanese]
\newblock Tokyo, Japan: Japan Statistical Association.

\bibitem{JGDC2008}
{Japan Geographic Data Center}. 2008 
 \emph{{Jyumin Kihon Dai-cho jinko yoran Heisei 20 nen ban}}.
\newblock [in Japanese]
\newblock Tokyo, Japan: Japan Statistical Association.

\bibitem{JGDC2009}
{Japan Geographic Data Center}. 2009 
 \emph{{Jyumin Kihon Dai-cho jinko yoran Heisei 20 nen ban}}.
\newblock [in Japanese]
\newblock Tokyo, Japan: Japan Statistical Association.

\bibitem{JGDC2010}
{Japan Geographic Data Center}. 2010 
 \emph{{Jyumin Kihon Dai-cho jinko yoran Heisei 21 nen ban}}.
\newblock [in Japanese]
\newblock Tokyo, Japan: Japan Statistical Association.

\end{thebibliography}
\end{document}